\documentclass{amsart}%
\usepackage{amsfonts}
\usepackage{amsmath}
\usepackage{amssymb}
\usepackage{graphicx}%
\providecommand{\U}[1]{\protect\rule{.1in}{.1in}}

\theoremstyle{plain}

\newtheorem{definition}{Definition}

\newtheorem{proposition}{Proposition}
\newtheorem{remark}{Remark}

\numberwithin{equation}{section}
\begin{document}
\title[QHS Algorithms]{Quantum Hidden Subgroup Algorithms: An Algorithmic Toolkit}
\author{Samuel J. Lomonaco, Jr.}
\address{University of Maryland Baltimore County (UMBC)\\
Baltimore, MD \ 21250}
\email{Lomonaco@umbc.edu}
\urladdr{http://www.csee.umbc.edu/\symbol{126}lomonaco}
\author{Louis H. Kauffman}
\curraddr{University of Illinois at Chicago (UIC)\\
Chicago, IL \ 60607}
\email{Kauffman@uic.edu}
\urladdr{http://www.math.uic.edu/\symbol{126}kauffman}
\date{May 22, 2006}

\begin{abstract}
One of the most promising and versatile approaches to creating new quantum
algorithms is based on the quantum hidden subgroup (QHS) paradigm, originally
suggested by Alexei Kitaev. \ This class of quantum algorithms encompasses the
Deutsch-Jozsa, Simon, Shor algorithms, and many more.

In this paper, our strategy for finding new quantum algorithms is to decompose
Shor's quantum factoring algorithm into its basic primitives, then to
generalize these primitives, and finally to show how to reassemble them into
new QHS algorithms. \ Taking an "alphabetic building blocks approach," we use
these primitives to form an "algorithmic toolkit" for the creation of new
quantum algorithms, such as wandering Shor algorithms, continuous Shor
algorithms, the quantum circle algorithm, the dual Shor algorithm, a QHS
algorithm for Feynman integrals, free QHS algorithms, and more. \ 

Toward the end of this paper, we show how Grover's algorithm is most
surprisingly \textquotedblleft almost\textquotedblright\ a QHS algorithm, and
how this result suggests the possibility of an even more complete "algorithmic
tookit" beyond the QHS algorithms.

\end{abstract}
\maketitle
\tableofcontents

\section{Introduction}

\bigskip

One major obstacle to the fulfillment of the promise of quantum computing is
the current scarcity of quantum algorithms. \ Quantum computing researchers
simply have not yet found enough quantum algorithms to determine whether or
not future quantum computers will be general purpose or special purpose
computing devices. \ As a result, much more research is crucially needed to
determine the algorithmic limits of quantum computing.

\bigskip

One of the most promising and versatile approaches to creating new quantum
algorithms is based on the quantum hidden subgroup (QHS) paradigm, originally
suggested by Alexei Kitaev \cite{Kitaev1}. \ This class of quantum algorithms
encompasses the Deutsch-Jozsa, Simon, Shor algorithms, and many more.\ 

\bigskip

In this paper, our strategy for finding new quantum algorithms is to decompose
Shor's quantum factoring algorithm into its basic primitives, then to
generalize these primitives, and finally to show how to reassemble them into
new QHS algorithms. \ Taking an "alphabetic building blocks approach," we will
use these primitives to form an "algorithmic toolkit" for the creation of new
quantum algorithms, such as wandering Shor algorithms, continuous Shor
algorithms, the quantum circle algorithm, the dual Shor algorithm, a QHS
algorithm for Feynman integrals, free QHS algorithms, and more.

\bigskip

Toward the end of this paper, we show how Grover's algorithm is most
surprisingly \textquotedblleft almost\textquotedblright\ a QHS algorithm, and
how this suggests the possibility of an even more complete "algorithmic
tookit" beyond the QHS algorithms.

\bigskip

\section{An example of Shor's quantum factoring algorithm}

\bigskip

Before discussing how Shor's algorithm can be decomposed into its primitive
components, let's take a quick look at an example of the execution of Shor's
factoring algorithm. \ As we discuss this example, we suggest that the reader,
as an exercise, try to find the basic QHS primitives that make up this
algorithm. \ Can you see them?

\bigskip

Shor's quantum factoring algorithm reduces the task of factoring a positive
integer $N$ to first finding a random integer $a$ relatively prime to $N$, and
then next to determining the period $P$ of the following function
\[%
\begin{array}
[c]{ccl}%
\mathbb{Z} & \overset{\varphi}{\longrightarrow} & \mathbb{Z}\operatorname{mod}%
N\\
x & \longmapsto & a^{x}\operatorname{mod}N\text{ ,}%
\end{array}
\]
where $\mathbb{Z}$ denotes the additive group of integers, and where
$\mathbb{Z}\operatorname{mod}N$ denotes the integers $\operatorname{mod}N$
under multiplication\footnote{A random integer $a$ with $\gcd\left(
a,N\right)  =1$ is found by selecting a random integer, and then applying the
Euclidean algorithm to determine whether or not it is relatively prime to $N$.
\ If not, then the $\gcd$ is a non-trivial factor of $N$, and there is no need
to proceed futher. \ However, this possibility is highly unlikely if $N$ is
large.}.

\bigskip

Since $\mathbb{Z}$ is an infinite group, Shor chooses to work instead with the
finite additive cyclic group $\mathbb{Z}_{Q}$ of order $Q=2^{m}$, where
$N^{2}\leq Q<2N^{2},$ and with the ``approximating'' map
\[%
\begin{array}
[c]{ccll}%
\mathbb{Z}_{Q} & \overset{\widetilde{\varphi}}{\longrightarrow} &
\mathbb{Z}\operatorname{mod}N & \\
x & \longmapsto & a^{x}\operatorname{mod}N\text{ ,} & 0\leq x<Q
\end{array}
\]

Shor begins by constructing a quantum system with two quantum registers
\[
\left\vert \text{\textsc{Left}\_\textsc{Register}}\right\rangle \left\vert
\text{\textsc{Right}\_\textsc{Register}}\right\rangle \text{ ,}%
\]
the left intended for holding the arguments $x$ of $\widetilde{\varphi}$, the
right for holding the corresponding values of $\widetilde{\varphi}$. \ This
quantum system has been constructed with a unitary transformation
\[
U_{\widetilde{\varphi}}:\left\vert x\right\rangle \left\vert 1\right\rangle
\longmapsto\left\vert x\right\rangle \left\vert \widetilde{\varphi}\left(
x\right)  \right\rangle
\]
implementing the \textquotedblleft approximating\textquotedblright\ map
$\widetilde{\varphi}$.

\bigskip

As an example, let us use Shor's algorithm to factor the\ integer $N=21$,
assuming that $a=2$ has been randomly chosen. \ Thus, $Q=2^{9}=512$.

\bigskip

Unknown to us, the period is $P=6$, and hence, $Q=6\cdot85+2$. \ 

\bigskip

We proceed by executing the following steps:

\bigskip

\begin{itemize}
\item[\fbox{$\mathbb{STEP}$\textbf{ 0}}] Initialize%
\[
\left|  \psi_{0}\right\rangle =\left|  0\right\rangle \left|  1\right\rangle
\]
\bigskip

\item[\fbox{$\mathbb{STEP}$\textbf{ 1}}] Apply the inverse Fourier
transform\footnote{Actually, for this step, the original Shor algorithm uses
instead the Hadamard transform, which for step 1, has the same effect as the
512-point Fourier transform.}
\[
\mathcal{F}^{-1}:\left\vert u\right\rangle \longmapsto\frac{1}{\sqrt{512}}%
\sum_{x=0}^{511}\omega^{-ux}\left\vert x\right\rangle
\]
to the left register, where $\omega=\exp(2\pi i/512)$ is a primitive $512$-th
root of unity, to obtain
\[
\left\vert \psi_{1}\right\rangle =\frac{1}{\sqrt{512}}\sum_{x=0}%
^{511}\left\vert x\right\rangle \left\vert 1\right\rangle
\]
\bigskip

\item[\fbox{$\mathbb{STEP}$\textbf{ 2}}] Apply the unitary transformation
\[
U_{\widetilde{\varphi}}:\left|  x\right\rangle \left|  1\right\rangle
\longmapsto\left|  x\right\rangle \left|  2^{x}\operatorname{mod}%
21\right\rangle
\]
to obtain
\[
\left|  \psi_{2}\right\rangle =\frac{1}{\sqrt{512}}\sum_{x=0}^{511}\left|
x\right\rangle \left|  2^{x}\operatorname{mod}21\right\rangle
\]
\bigskip

\item[\fbox{$\mathbb{STEP}$\textbf{ 3}}] Apply the Fourier transform
\[
\mathcal{F}:\left\vert x\right\rangle \longmapsto\frac{1}{\sqrt{512}}%
\sum_{y=0}^{511}\omega^{xy}\left\vert y\right\rangle
\]
to the left register to obtain
\begin{align*}
\left\vert \psi_{3}\right\rangle  &  =\frac{1}{512}\sum_{x=0}^{511}\sum
_{y=0}^{511}\omega^{xy}\left\vert y\right\rangle \left\vert 2^{x}%
\operatorname{mod}21\right\rangle =\frac{1}{512}\sum_{y=0}^{511}\left\vert
y\right\rangle \left(  \sum_{x=0}^{511}\omega^{xy}\left\vert 2^{x}%
\operatorname{mod}21\right\rangle \right) \\
& \\
&  =\frac{1}{512}\sum_{y=0}^{511}\left\vert y\right\rangle \left\vert
\Upsilon\left(  y\right)  \right\rangle
\end{align*}
where
\[
\left\vert \Upsilon\left(  y\right)  \right\rangle =\sum_{x=0}^{511}%
\omega^{xy}\left\vert 2^{x}\operatorname{mod}21\right\rangle
\]
\bigskip

\item[\fbox{$\mathbb{STEP}$\textbf{ 4}}] Measure the left register. \ Then
with Probability
\[
Prob_{\widetilde{\varphi}}\left(  y\right)  =\frac{\left\langle \ \Upsilon
\left(  y\right)  \mid\Upsilon\left(  y\right)  \ \right\rangle }{\left(
512\right)  ^{2}}%
\]
the state will ``collapse'' to $\left|  y\right\rangle $ with the value
measured being the integer $y$, where $0\leq y<Q$.
\end{itemize}

\bigskip

A plot of $Prob_{\widetilde{\varphi}}\left(  y\right)  $ is shown in Figure 1.
\ (See \cite{Lomonaco2} and \cite{Lomonaco4} for details.)%
\begin{center}
\includegraphics[
height=3.4523in,
width=4.3543in
]%
{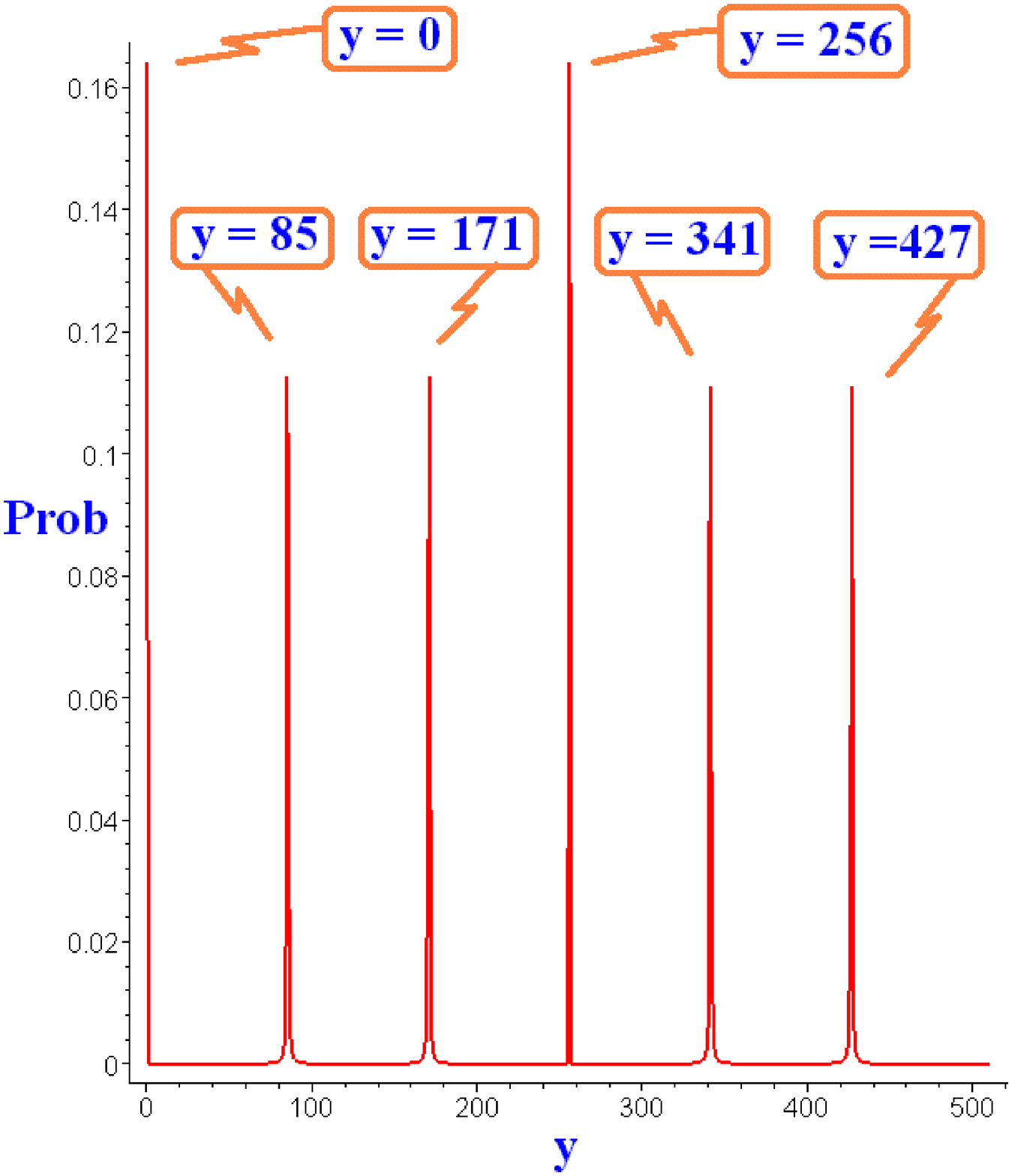}%
\\
\textbf{Figure 1. \ A plot of }$\mathbf{Prob}_{\widetilde{\mathbf{\varphi}}%
}\mathbf{(y)}$\textbf{.}%
\end{center}

\bigskip

The peaks in the above plot of $Prob_{\widetilde{\varphi}}\left(  y\right)  $
occur at the integers
\[
y=0,\ 85,\ 171,\ 256,\ 341,\ 427\text{.}%
\]
The probability that at least one of these six integers will occur is quite
high. \ It is actually $0.78^{+}$. \ Indeed, the probability distribution has
been intentionally engineered to make the probability of these particular
integers as high as possible. \ And there is a good reason for doing so.

\bigskip

The above six integers are those for which the corresponding rational $y/Q$ is
\textquotedblleft closest\textquotedblright\ to a rational of the form $d/P$.
\ By \textquotedblleft closest\textquotedblright\ we mean that
\[
\left\vert \frac{y}{Q}-\frac{d}{P}\right\vert <\frac{1}{2Q}<\frac{1}{2P^{2}%
}\text{ .}%
\]
In particular,
\[
\frac{0}{512},\ \frac{85}{512},\ \frac{171}{512},\frac{256}{512},\ \frac
{341}{512},\ \frac{427}{512}%
\]
are rationals respectively \textquotedblleft closest\textquotedblright\ to the
rationals
\[
\frac{0}{6},\ \frac{1}{6},\ \frac{2}{6},\ \frac{3}{6},\ \frac{4}{6},\ \frac
{5}{6}\text{ .}%
\]
The six rational numbers $0/6,\ 1/6,\ \ldots\ ,\ 5/6$ are "closest" in the
sense that they are convergents of the continued fraction expansions of
$0/512,\ 85/512,\ \ldots\ ,\ 427/512$, respectively. \ Hence, each of the six
rationals $0/6,\ 1/6,\ \ldots\ ,\ 5/6$ can be found using the standard
continued fraction recursion formulas.

\bigskip

But ... , we are not searching for rationals of the form $d/P$. \ Instead, we
seek only the denominator $P=6$. \ 

\bigskip

Unfortunately, the denominator $P=6$ can only be obtained from the continued
fraction recursion when the numerator and denominator of $d/P$ are relatively
prime. \ Given that the algorithm has selected one of the random integers
$0,\ 85,\ \ldots\ ,\ 427$, the probability that the corresponding rational
$d/P$ has relatively prime numerator and denominator is $\phi\left(  6\right)
/6=1/3$, where $\phi\left(  -\right)  $ denotes the Euler phi (totient)
function. \ So the probability of finding $P=6$ is actually not $0.78^{+}$,
but is instead $0.23^{-}$.

\bigskip

As it turns out, if he repeats the algorithm $O\left(  \lg\lg N\right)  $
times, we will obtain the desired period $P$ with probability bounded below by
approximately $4/\pi^{2}$. \ However, this is not the end of the story. \ Once
we have in our possession a candidate $P^{\prime}$ for the actual period
$P=6$, the only way we can be sure we have the correct period $P$ is to test
$P^{\prime}$ by computing $2^{P^{\prime}}\operatorname{mod}21$. \ If the
result is $1$, we are certain we have found the correct period $P$. \ This
last part of the computation is done by the repeated squaring
algorithm\footnote{By the repeated squaring algorithm, we mean the algorithm
which computes $a^{P^{\prime}}\operatorname{mod}N$ via the expression
\[
a^{P^{\prime}}=\prod_{j}\left(  a^{2^{j}}\right)  ^{P_{j}^{\prime}}\text{,}%
\]
where $P^{\prime}=\sum_{j}P_{j}^{\prime}2^{j}$ is the radix 2 expansion of
$P^{\prime}$.}.

\bigskip

\section{Definition of quantum hidden subgroup (QHS) algorithms}

\bigskip

Now that we have taken a quick look at Shor's algorithm, let's see how it can
be decomposed into its primitive algorithmic components. \ We will first need
to answer the following question:

\begin{center}
\textit{What is a quantum hidden subgroup algorithm?}
\end{center}

\bigskip

But before we can answer the this question, we need to provide an answer to an
even more fundamental question:

\bigskip

\begin{center}
\textit{What is a hidden subgroup problem?}
\end{center}

\bigskip

\begin{definition}
A map $\varphi:G\longrightarrow S$ from a group $G$ into a set $S$ is said to
have \textbf{hidden subgroup structure} if there exists a subgroup
$K_{\mathbf{\varphi}}$ of $G$, called a \textbf{hidden subgroup}, and an
injection $\iota_{\mathbf{\varphi}}:G/K_{\mathbf{\varphi}}\longrightarrow S$,
called a \textbf{hidden injection}, such that the diagram%
\[%
\begin{array}
[c]{ccc}%
G & \overset{\varphi}{\longrightarrow} & S\\
\nu\searrow &  & \nearrow\iota_{\mathbf{\varphi}}\\
& G/K_{\mathbf{\varphi}} &
\end{array}
\]
is commutative\footnote{By saying that this diagram is commutative, we mean
$\varphi=\iota_{\varphi}\circ\nu$. \ The notion generalizes in an obvious way
to more complicated diagrams.}, where $G/K_{\varphi}$ denotes the collection
of right cosets of $K_{\varphi}$ in $G$, and where $\nu:G\longrightarrow
G/K_{\varphi}$ is the natural surjection of $G$ onto $G/K_{\mathbf{\varphi}}$.
\ We refer to the group $G$ as the \textbf{ambient group} and to the set $S$
as the \textbf{target set}. \ If $K_{\mathbf{\varphi}}$ is a normal subgroup
of $G$, then $H_{\mathbf{\varphi}}=G/K_{\mathbf{\varphi}}$ is a group, called
the \textbf{hidden quotient group}, and $\nu:G\longrightarrow G/K_{\varphi}$
is an epimorphism, called the \textbf{hidden epimorphism}. We will call the
above diagram the \textbf{hidden subgroup structure} of the map $\varphi
:G\longrightarrow S$. (See \cite{Lomonaco4},\cite{Kitaev1}.)
\end{definition}

\bigskip

\begin{remark}
The underlying intuition motivating this formal definition is as follows:
Given a natural surjection (or epimorphism) $\nu:G\longrightarrow
G/K_{\varphi}$, an "archvillain with malice of forethought" hides the
algebraic structure of $\nu$ by intentionally renaming all the elements of
$G/K_{\mathbf{\varphi}}$ , and "maliciously tossing in for good measure" some
extra elements to form a set $S$ and a map $\varphi:G\longrightarrow S$.
\end{remark}

\bigskip

The hidden subgroup problem can be stated as follows:

\bigskip

\noindent\textbf{Hidden Subgroup Problem (HSP).} \textit{Let }$\varphi
:G\longrightarrow S$\textit{ be a map with hidden subgroup structure. The
problem of determining a hidden subgroup }$K_{\varphi}$\textit{ of }%
$G$\textit{ is called a \textbf{hidden subgroup problem (HSP)}. \ An algorithm
solving this problem is called a \textbf{hidden subgroup algorithm}.}

\bigskip

The corresponding quantum form of this HSP is stated as follows:

\bigskip

\noindent\textbf{Hidden Subgroup Problem (Quantum Version).} \textit{Let
}$\varphi:G\longrightarrow S$\textit{ be a map with hidden subgroup structure.
\ Construct a quantum implementation of the map }$\varphi$\textit{ as follows:
\newline Let }$\mathcal{H}_{G}$\textit{ and }$\mathcal{H}_{S}$\textit{ be
Hilbert spaces defined respectively by the orthonormal bases }$\left\{
\left\vert g\right\rangle :g\in G\right\}  $\textit{and }$\left\{  \left\vert
s\right\rangle :s\in S\right\}  $ \textit{and let }$s_{0}=\varphi\left(
1\right)  $\textit{, where }$1$\textit{ denotes the identity\footnote{We are
using multiplicative notation for the group $G$.} of the ambient group }%
$A$\textit{. \ Finally, let }$U_{\varphi}$\textit{ be a unitary transformation
such that}%
\[%
\begin{array}
[c]{ccc}%
\mathcal{H}_{G}\otimes\mathcal{H}_{S} & \longrightarrow & \mathcal{H}%
_{G}\otimes\mathcal{H}_{S}\\
\left\vert g\right\rangle \left\vert s_{0}\right\rangle  & \mapsto &
\left\vert g\right\rangle \left\vert \varphi(g)\right\rangle
\end{array}
\]
\textit{Determine the hidden subgroup }$K_{\varphi}$\textit{ with bounded
probability of error by making as few queries as possible to the blackbox
}$U_{\mathbf{\varphi}}$\textit{. \ A quantum algorithm solving this problem is
called a \textbf{quantum hidden subgroup (QHS) algorithm}.}

\bigskip

\section{The generic QHS algorithm}

\bigskip

We are now in a position to construct one of the fundamental algorithmic
primitives found in Shor's algorithm.

\bigskip

Let $\varphi:G\longrightarrow S$ be a map from a group $G$ to a set $S$ with
hidden subgroup structure. We assume that all representations of $G$ are
equivalent to unitary representations\footnote{This is true for all finite
groups as well as for a large class of infinite groups.}. \ Let $\widehat{G}$
denote a \textbf{complete set of distinct irreducible unitary representations}
of $G$. \ Using multiplicative notation for $G$, we let $1$ denote the
\textbf{identity} of $G$, and let $s_{0}$ denote its image in $S$. Finally,
let $\widehat{1}$ denote the \textbf{trivial representation} of $G$.

\begin{remark}
If $G$ is abelian, then $\widehat{\mathbf{G}}$ becomes the \textbf{dual group}
of characters.
\end{remark}

\bigskip

The generic QHS algorithm is given below:

\bigskip

\begin{center}
Quantum Subroutine \textsc{QRand}$\left(  \varphi\right)  $
\end{center}

\bigskip

\begin{itemize}
\item[\textbf{Step 0.}] Initialization%
\[
\left\vert \psi_{0}\right\rangle =\left\vert \widehat{1}\right\rangle
\left\vert s_{\mathbf{0}}\right\rangle \in\mathcal{H}_{\widehat{G}}%
\otimes\mathcal{H}_{S}%
\]

\item[\textbf{Step 1.}] Application of the inverse Fourier transform
$\mathcal{F}_{G}^{-1}$of $G$ to the left register%
\[
\left\vert \psi_{1}\right\rangle =\dfrac{1}{\sqrt{\left\vert G\right\vert }%
}\sum\limits_{g\in G}\left\vert g\right\rangle \left\vert s_{0}\right\rangle
\in\mathcal{H}_{G}\otimes\mathcal{H}_{S}\text{ \ ,}%
\]
where $\left\vert G\right\vert $ denotes the cardinality of the group
$G$.\bigskip

\item[\textbf{Step 2.}] Application of the unitary transformation $U_{\varphi
}$%
\[
\left\vert \psi_{2}\right\rangle =\dfrac{1}{\sqrt{\left\vert G\right\vert }%
}\sum\limits_{g\in G}\left\vert g\right\rangle \left\vert \varphi
(g)\right\rangle \in\mathcal{H}_{G}\otimes\mathcal{H}_{S}%
\]
\bigskip

\item[\textbf{Step 3.}] Application of the Fourier transform $\mathcal{F}_{G}$
of $G$ to the left register%
\[
\left\vert \psi_{3}\right\rangle =\dfrac{1}{\left\vert G\right\vert }%
\sum\limits_{\gamma\in\widehat{G}}\left\vert \gamma\right\vert Trace\left(
\sum\limits_{\mathbf{g}\mathbf{\in}\mathbf{G}}\gamma^{\dag}\left(  g\right)
\left\vert \gamma\right\rangle \right)  \left\vert \varphi(g)\right\rangle
=\dfrac{1}{\left\vert G\right\vert }\sum\limits_{\gamma\in\widehat{G}%
}\left\vert \gamma\right\vert Trace\left(  \left\vert \gamma\right\rangle
\left\vert \Phi(\gamma^{\mathbf{\dag}})\right\rangle \right)  \in
\mathcal{H}_{\widehat{G}}\otimes\mathcal{H}_{S}\text{ \ ,}%
\]
where $\left\vert \mathbf{\gamma}\right\vert $ denotes the degree of the
representation $\gamma$, where $\gamma^{\mathbf{\dag}}$ denotes the
contragradient representation (i.e., $\gamma^{\mathbf{\dag}}\left(  g\right)
=\gamma\left(  g^{-1}\right)  ^{T}=\overline{\gamma\left(  g\right)  }^{T}$),
where $Trace\left(  \gamma^{\mathbf{\dag}}\left\vert \gamma\right\rangle
\right)  =\sum\limits_{i=1}^{\left\vert \gamma\right\vert }\sum\limits_{j=1}%
^{\left\vert \gamma\right\vert }\overline{\gamma_{ji}\left(  g\right)
}\left\vert \gamma_{ij}\right\rangle $, and where $\left\vert \Phi\left(
\gamma_{ij}^{\dag}\right)  \right\rangle =\sum\limits_{g\in G}\overline
{\gamma_{ji}\left(  g\right)  }\left\vert \varphi\left(  g\right)
\right\rangle $.\bigskip

\item[\textbf{Step 4.}] Measurement of the left quantum register with respect
to the orthonormal basis%
\[
\left\{  \left\vert \gamma_{ij}\right\rangle :\gamma\in\widehat{G},\,1\leq
i,j\leq\left\vert \gamma\right\vert \right\}  \text{ \ .}%
\]
Thus, with probability%
\[
Prob_{\varphi}\left(  \gamma_{ij}\right)  =\dfrac{\left\vert \gamma\right\vert
^{2}\left\langle \Phi\left(  \gamma_{ij}^{\dag}\right)  |\Phi\left(
\gamma_{ij}^{\dag}\right)  \right\rangle }{\left\vert G\right\vert ^{2}}%
\]
the resulting measured value is the entry $\gamma_{\mathbf{i}\mathbf{j}}$, and
the quantum system "collapses" to the state%
\[
\left\vert \psi_{4}\right\rangle =\dfrac{\left\vert \gamma_{ij}\right\rangle
\left\vert \Phi\left(  \gamma_{ij}^{\dag}\right)  \right\rangle }%
{\sqrt{\left\langle \Phi\left(  \gamma_{ij}^{\dag}\right)  |\Phi\left(
\gamma_{ij}^{\dag}\right)  \right\rangle }}\in\mathcal{H}_{\widehat{G}}%
\otimes\mathcal{H}_{S}%
\]
\bigskip

\item[\textbf{Step 5.}] Step 5. Output $\gamma_{ij}$, and stop.
\end{itemize}

\bigskip

\section{Pushing and Lifting hidden subgroup problems (HSPs)}

\bigskip

But Shor's algorithm consists of more than the primitive \textsc{QRand}.

\bigskip

For many (but not all) hidden subgroup problems (HSPs) $\varphi
:G\longrightarrow S$, the corresponding generic QHS algorithm \textsc{QRand}
either is not physically implementable or is too expensive to implement
physically. \ For example, the HSP $\varphi$ is usually not physically
implementable if the ambient group is infinite (e.g., $G$ is the infinite
cyclic group $\mathbb{Z}$), and is too expensive to implement if the ambient
group is too large (e.g., $G$ is the symmetric group $\mathbb{S}_{10^{100}}$).
\ In this case, there is a standard generic way of "tweaking" the HSP to get
around this problem, which we will call \textbf{pushing}.

\bigskip

\begin{definition}
Let $\varphi:G\longrightarrow S$ be a map from a group $G$ to a set $S$. \ A
map $\widetilde{\varphi}:\widetilde{G}\longrightarrow S$ from a group
$\widetilde{G}$ to the set $S$ is said to be a \textbf{push} of $\varphi$,
written
\[
\widetilde{\varphi}=Push\left(  \varphi\right)  \text{ \ ,}%
\]
provided there exists an epimorphism $\nu:G\longrightarrow\widetilde{G}$ from
$G$ onto $\widetilde{G}$, and a transversal\footnote{Let $\nu:A\longrightarrow
B$ be an epimorphism from a group $A$ to a group $B$. \ Then a transversal
$\tau$ of $\nu$ is is a map $\tau:B\longrightarrow A$ such that $\nu\circ
\tau:B\longrightarrow A$ is the identity map $b\longmapsto b$. \ (It
immediately follows that $\tau$ is an injection.) \ In othere words, a
transversal $\tau$ of an epimorphism $\nu$ is a map which maps eacl element
$b$ of $B$ to an element of $A$ contained in the coset $b$, i.e., to a coset
representative of $b$.} $\tau:\widetilde{G}\longrightarrow G$ of $\nu$ such
that $\widetilde{\varphi}=\varphi\circ\tau$, i.e., such that the following
diagram is commutative%
\[%
\begin{array}
[c]{ccc}%
G & \overset{\varphi}{\longrightarrow}\quad & S\\
\quad\uparrow\tau & \quad\nearrow\widetilde{\varphi}\quad & \\
\widetilde{G} &  &
\end{array}
\]

\end{definition}

\bigskip

If the epimorphism $\mu$ and the transversal $\tau$ are chosen in an
appropriate way, then execution of the generic QHS subroutine with input
$\widetilde{\varphi}=Push\left(  \varphi\right)  $ , i.e., execution of
\[
\text{\textsc{QRand}}\left(  \widetilde{\varphi}\right)  \text{ \ ,}%
\]
will with high probability produce an irreducible representation
$\widetilde{\gamma}$ of the group $\widetilde{G}$ which is sufficiently close
to an irreducible representation $\gamma$ of the group $G$. \ If this is the
case, then there is a polynomial time classical algorithm which upon input
$\widetilde{\gamma}$ produces the representation $\gamma$. \ 

\bigskip

Obviously, much more can be said about pushing. \ But unfortunately that would
take us far afield from the objectives of this paper. \ For more information
on pushing, we refer the reader to \cite{Lomonaco7}.

\bigskip

It would be amiss not to mention that the above algorithmic primitive of
pushing suggests the definition of a second primitive which we will call
\textbf{lifting}.

\bigskip

\begin{definition}
Let $\varphi:G\longrightarrow S$ be a map from a group $G$ to a set $S$. \ A
map \underline{$\varphi$}$:\underline{G}\longrightarrow S$ from a group
$\underline{G}$ to the set $S$ is said to be a \textbf{lift} of $\varphi$,
written
\[
\underline{\varphi}=Lift\left(  \varphi\right)  \text{ \ ,}%
\]
provided there exists an morphism $\eta:\underline{G}\longrightarrow G$ from
\underline{$G$} to $G$ such that $\underline{\varphi}=\varphi\circ\eta$, i.e.,
such that the following diagram is commutative%
\[%
\begin{array}
[c]{cl}%
\underline{G} & \\
\ \ \eta\downarrow\quad & \quad\searrow\underline{\varphi}\quad\\
G & \quad\overset{\varphi}{\longrightarrow}\quad S
\end{array}
\]

\end{definition}

\bigskip%

\begin{center}
\includegraphics[
natheight=7.499600in,
natwidth=9.999800in,
height=2.578in,
width=3.4281in
]%
{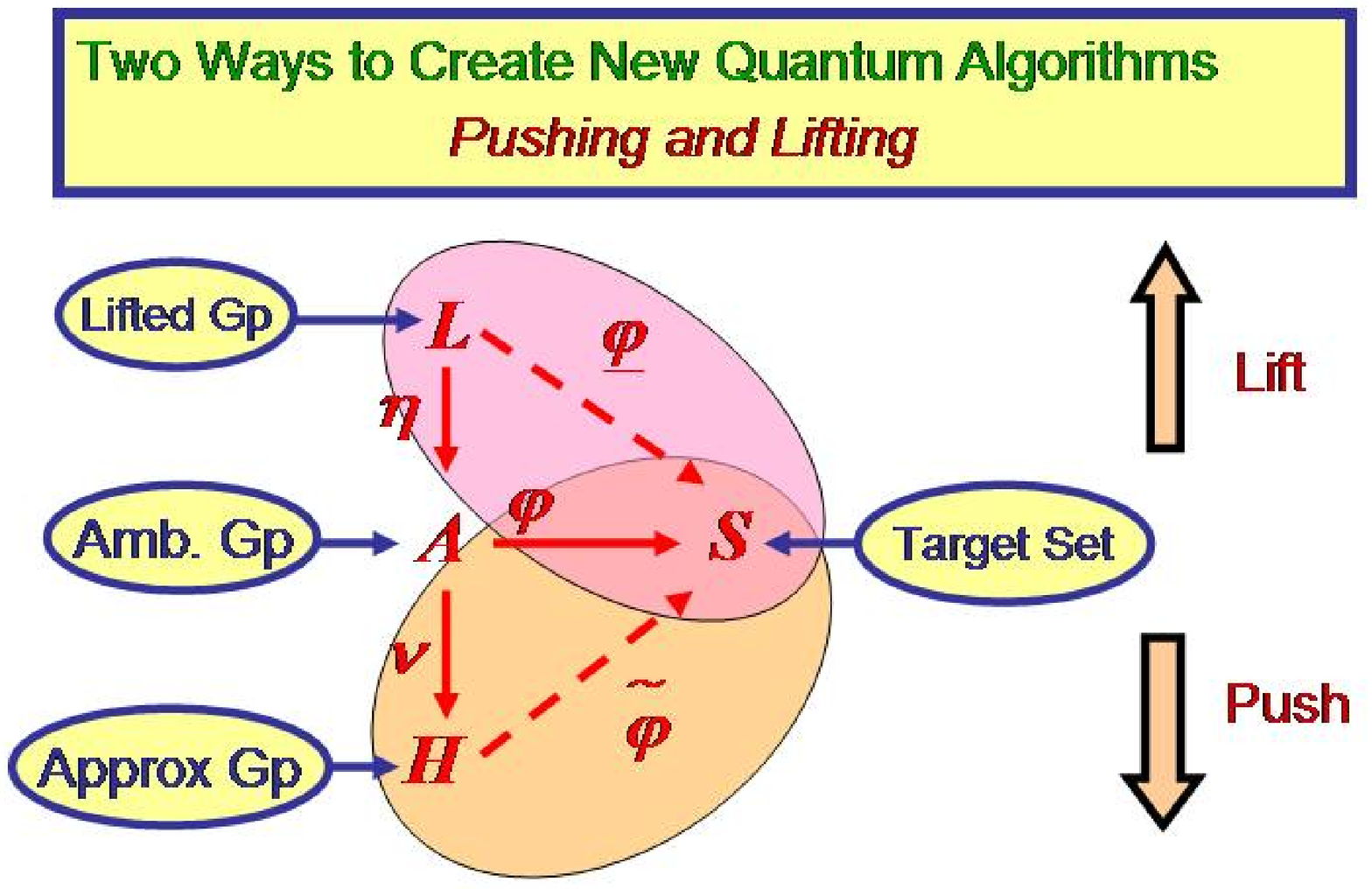}%
\\
\textbf{Figure 2. \ Pushing and Lifting HSPs.}%
\end{center}

\bigskip

\section{Shor's algorithm revisited}

\bigskip

We are now in position to describe Shor's algorithm in terms of its primitive
components. \ In particular, we are now in a position to see that Shor's
factoring algorithm is a classic example of a QHS algorithm created from the
push of an HSP.

\bigskip

Let $N$ be the integer to be factored. \ Let $\mathbb{Z}$ denote the additive
group of integers, and $\mathbb{Z}_{N}^{\times}$ denote the integers
$\operatorname{mod}N$ under multiplication. \ 

\bigskip

Shor's algorithm is a QHS algorithm that solves the following HSP
\[%
\begin{array}
[c]{rrc}%
\varphi:\mathbb{Z} & \longrightarrow & \mathbb{Z}_{N}^{\times}\\
m & \longmapsto & a^{m}\operatorname{mod}N
\end{array}
\]
with unknown hidden subgroup structure given by the following commutative
diagram
\[%
\begin{array}
[c]{ccc}%
\mathbb{Z} & \overset{\varphi}{\longrightarrow} & \mathbb{Z}_{N}^{\times}\\
\nu\searrow &  & \nearrow\iota\\
& \mathbb{Z}/P\mathbb{Z} &
\end{array}
\text{ \ ,}%
\]
where $a$ is an integer relatively prime to $N$, where $P$ is the hidden
integer period of the map $\varphi:\mathbb{Z}\longrightarrow\mathbb{Z}%
_{N}^{\times}$, where $P\mathbb{Z}$ is the additive subgroup of all integer
multiples of $P$ (i.e., the hidden subgroup), where $\nu:\mathbb{Z}%
\longrightarrow\mathbb{Z}/P\mathbb{Z}$\ is the natural epimorpism of the
integers onto the quotient group $\mathbb{Z}/P\mathbb{Z}$ (i.e., the hidden
epimorphism), and where $\iota:\mathbb{Z}/P\mathbb{Z\longrightarrow Z}%
_{N}^{\times}$ is the hidden monomorphism.

\bigskip

An obstacle to creating a physically implementable algorithm for this HSP is
that the domain $\mathbb{Z}$ of $\varphi$ is infinite. \ As observed by Shor,
a way to work around this difficulty is to push the HSP.

\bigskip

In particular, as illustrated by the following commutative diagram%
\[%
\begin{array}
[c]{ccl}%
\mathbb{Z\qquad} & \overset{\varphi}{\longrightarrow} & \qquad\mathbb{Z}%
_{N}^{\times}\\
\mu\searrow\nwarrow\tau &  & \nearrow\varphi=Push\left(  \varphi\right)
=\varphi\circ\tau\\
& \mathbb{Z}_{Q} &
\end{array}
\text{ \ \ ,}%
\]
a push $\widetilde{\varphi}=Push\left(  \varphi\right)  $ is constructed by
selecting the epimorphism $\mu:\mathbb{Z\longrightarrow Z}_{Q}$ of
$\mathbb{Z}$ onto the finite cyclic group $\mathbb{Z}_{Q}$ of order $Q$, where
the integer $Q$ is the unique power of $2$ such that $N^{2}\leq Q<2N^{2}$, and
then choosing the transversal\footnote{A \textbf{transversal} for an
epimorphism $\alpha_{\varphi}:\mathbb{Z\longrightarrow Z}_{Q}$ is an injection
$\tau_{\varphi}:\mathbb{Z_{Q}\longrightarrow Z}$ such that $\alpha_{\varphi
}\circ\tau_{\varphi}$ is the identity map on $\mathbb{Z}_{Q}$, i.e., a map
that takes each element of $\mathbb{Z}_{Q}$ onto a coset representative of the
element in $\mathbb{Z}$ .}
\[%
\begin{array}
[c]{rrc}%
\tau:\mathbb{Z}_{Q} & \longrightarrow & \mathbb{Z}\\
m\operatorname{mod}Q & \longmapsto & m
\end{array}
\text{ \ ,}%
\]
where $0\leq m<Q$. \ \textit{This push} $\widetilde{\varphi}=Push\left(
\varphi\right)  $ \textit{is called} \textbf{Shor's oracle}.

\bigskip

Shor's algorithm consists in first executing the quantum subroutine
\textsc{QRand}$\left(  \widetilde{\varphi}\right)  $, thereby producing a
random character
\[
\gamma_{y/Q}:m\operatorname{mod}Q\longmapsto\frac{my}{Q}\operatorname{mod}1
\]
of the finite cyclic group $\mathbb{Z}_{Q}$. \ The transversal $\tau$ used in
pushing has been engineered in such a way as to assure that the character
$\gamma_{y/Q}$ is sufficiently close to a character
\[
\gamma_{d/P}:k\operatorname{mod}P\longmapsto\frac{kd}{P}\operatorname{mod}1
\]
of the hidden quotient group $\mathbb{Z}/P\mathbb{Z}=\mathbb{Z}_{P}$. \ In
this case "sufficiently close" means that
\[
\left\vert \frac{y}{Q}-\frac{d}{P}\right\vert \leq\frac{1}{2P^{2}}\text{ \ ,}%
\]
which means that $d/P$ is a continued fraction convergent of $y/Q$, and thus
can be found found by the classical polynomial time continued fraction
algorithm\footnote{The characters $\gamma_{y/Q}$ and $\gamma_{d/P}$ can in the
obvious way be identified with points of in the unit circle in the complex
plane. \ With this identification, we can see that this inequalty is
equivalent to saying the the chordal distance bewteen these two rational
points on the unit circle is less than or equan to $1/2P^{2}$. \ Hence, Shor's
algorithm is using the topology of the unit circle.}.

\bigskip

\section{Wandering Shor algorithms, a.k.a., vintage Shor algorithms.}

\bigskip

Now let's use the primitives described in sections 3, 4, and 5 to create other
new QHS algorithms, called wandering Shor algorithms.

\bigskip

Wandering Shor algorithms are essentially QHS algorithms on free abelian
finite rank $n$ groups $A$ which, with each iteration, first select a random
cyclic direct summand $\mathbb{Z}$ of the group $A$, and then apply one
iteration of the standard Shor algorithm to produce a random character of the
\textquotedblleft approximating\textquotedblright\ finite group $\widetilde
{A}=\mathbb{Z}_{Q}$, called a \textbf{group probe}\footnote{By a group probe
$\widetilde{A}$, we mean an epimorphic image of the ambient group $A$.}. Three
different wandering Shor algorithms are created in \cite{Lomonaco4}. The first
two wandering Shor algorithms given in \cite{Lomonaco4} are quantum algorithms
which find the order $P$ of a maximal cyclic subgroup of the hidden quotient
group $H_{\varphi}$. The third computes the entire hidden quotient group
$H_{\varphi}$.

\bigskip

The first step in creating a wandering Shor algorithm is to find the right
generalization one of the primitives found in Shor's algorithm, namely, the
transversal $\iota:\mathbb{Z}_{Q}\longrightarrow\mathbb{Z}$ of Shor's
factoring algorithm. \ In other words, we need to construct the "correct"
generalization of the transversal from $\mathbb{Z}_{Q}$ to a free abelian
group $A$ of rank $n$. \ For this reason, we have created the following definition:

\bigskip

\begin{definition}
Let $A$ be the free abelian group of rank $n$, let $\nu:A\longrightarrow
\mathbb{Z}_{Q}$ onto the cyclic group $\mathbb{Z}_{Q}$ of order $Q$ with
selected generator $\widetilde{a}$. \ A transversal\footnote{Let
$\nu:A\longrightarrow B$ be an epimorphism from a group $A$ to a group $B$.
\ Then a transversal $\tau$ of $\nu$ is is a map $\tau:B\longrightarrow A$
such that $\nu\circ\tau:B\longrightarrow A$ is the identity map $b\longmapsto
b$. \ (It immediately follows that $\tau$ is an injection.) \ In othere words,
a transversal $\tau$ of an epimorphism $\nu$ is a map which maps eacl element
$b$ of $B$ to an element of $A$ contained in the coset $b$, i.e., to a coset
representative of $b$.} $\iota:\mathbb{Z}_{Q}\longrightarrow A$ of $\nu$ is
said to be a \textbf{Shor transversal} provided that:

\begin{itemize}
\item[\textbf{1)}] $\iota\left(  n\widetilde{a}\right)  =n\iota\left(
\widetilde{a}\right)  $ for all $0\leq n<Q$

\item[\textbf{2)}] For each (free abelian) basis $a_{1}^{\prime},a_{2}%
^{\prime},\ldots,a_{n}^{\prime}$ of $A$, the coefficients $\lambda_{1}%
^{\prime},\lambda_{2}^{\prime},\ldots,\lambda_{n}^{\prime}$ of $\iota\left(
\widetilde{a}\right)  =\sum\nolimits_{j}\lambda_{j}^{\prime}a_{j}^{\prime}$
satisfy $\gcd\left(  \lambda_{1}^{\prime},\lambda_{2}^{\prime},\ldots
,\lambda_{n}^{\prime}\right)  =1$.
\end{itemize}
\end{definition}

\bigskip

\begin{remark}
Later, when we construct a generalization of Shor transversals to free groups
of finite rank $n$, we will see that the first condition simply states that a
Shor transversal is nothing more than a 2-sided Schreier transversal. \ The
second condition of the above definition simply says that $\iota$ maps the
generator $\widetilde{a}$ of $\mathbb{Z}_{Q}$ onto a generator of a free
direct summand $\mathbb{Z}$ of $A$. (For more details, please refer to section
12 of this paper.)
\end{remark}

\bigskip

\begin{remark}
In \cite{Lomonaco4}, we show how to use the extended Euclidean algorithm to
construct the epimorphism $\nu:A\longrightarrow\mathbb{Z}_{Q}$ and the
transversal $\iota:\mathbb{Z}_{Q}\longrightarrow A$.
\end{remark}

\bigskip%

\begin{center}
\includegraphics[
height=5.3973in,
width=3.2543in
]%
{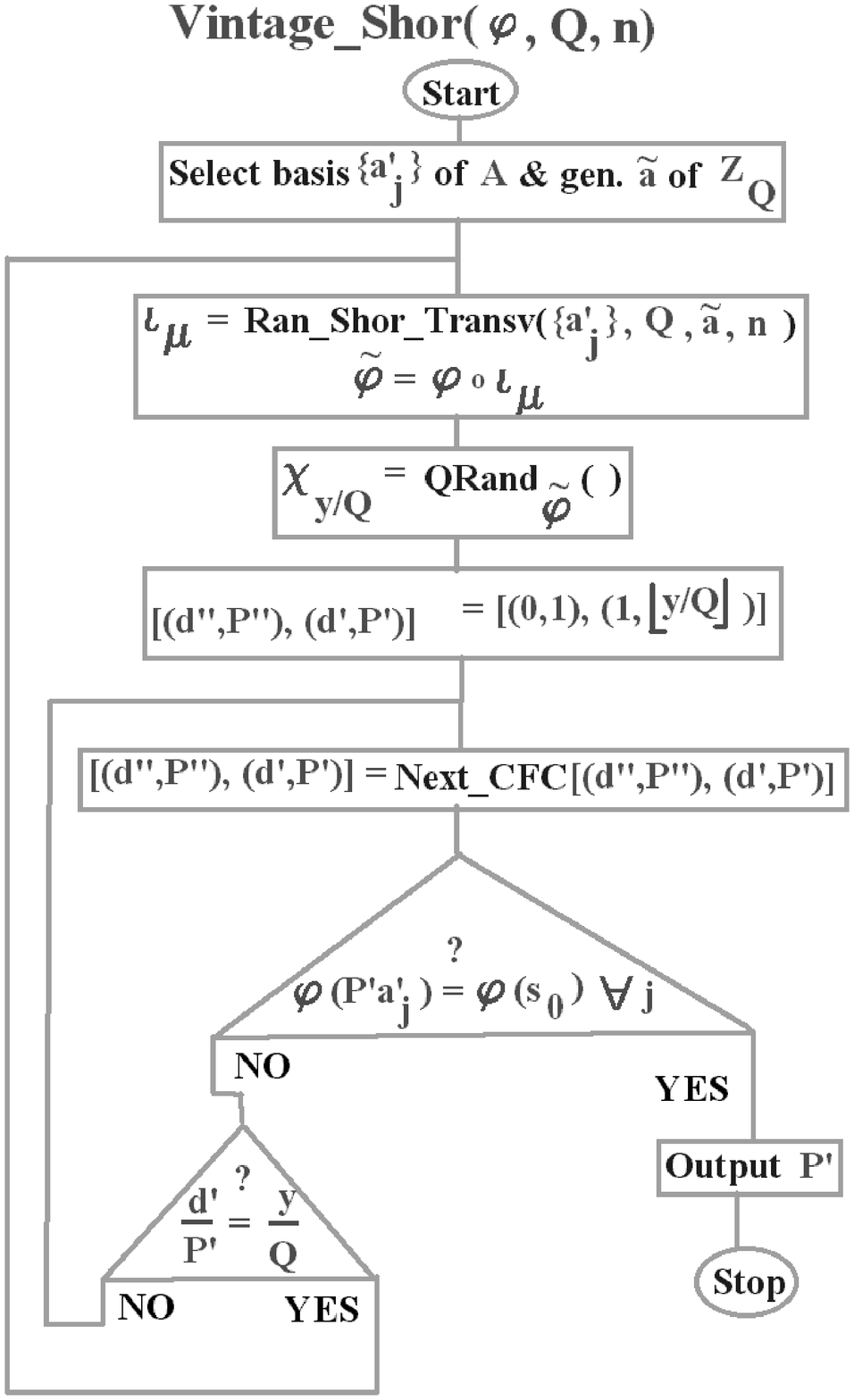}%
\\
\textbf{Figure 3. Flowchart for the first wandering Shor algorithm (a.k.a., a
vintage Shor algorithm). \ This algorithm finds the order }$P$\textbf{\ of a
maximal cyclic subgroup of the hidden quotient group }$H_{\varphi}$\textbf{.}%
\end{center}

\bigskip

Flow charts for the three wandering Shor algorithms created in
\cite{Lomonaco4} are given in figures 3 through 5. In \cite{Lomonaco4}, these
were also called \textbf{vintage Shor algorithms}.

\bigskip%

\begin{center}
\includegraphics[
height=6.6694in,
width=3.8882in
]%
{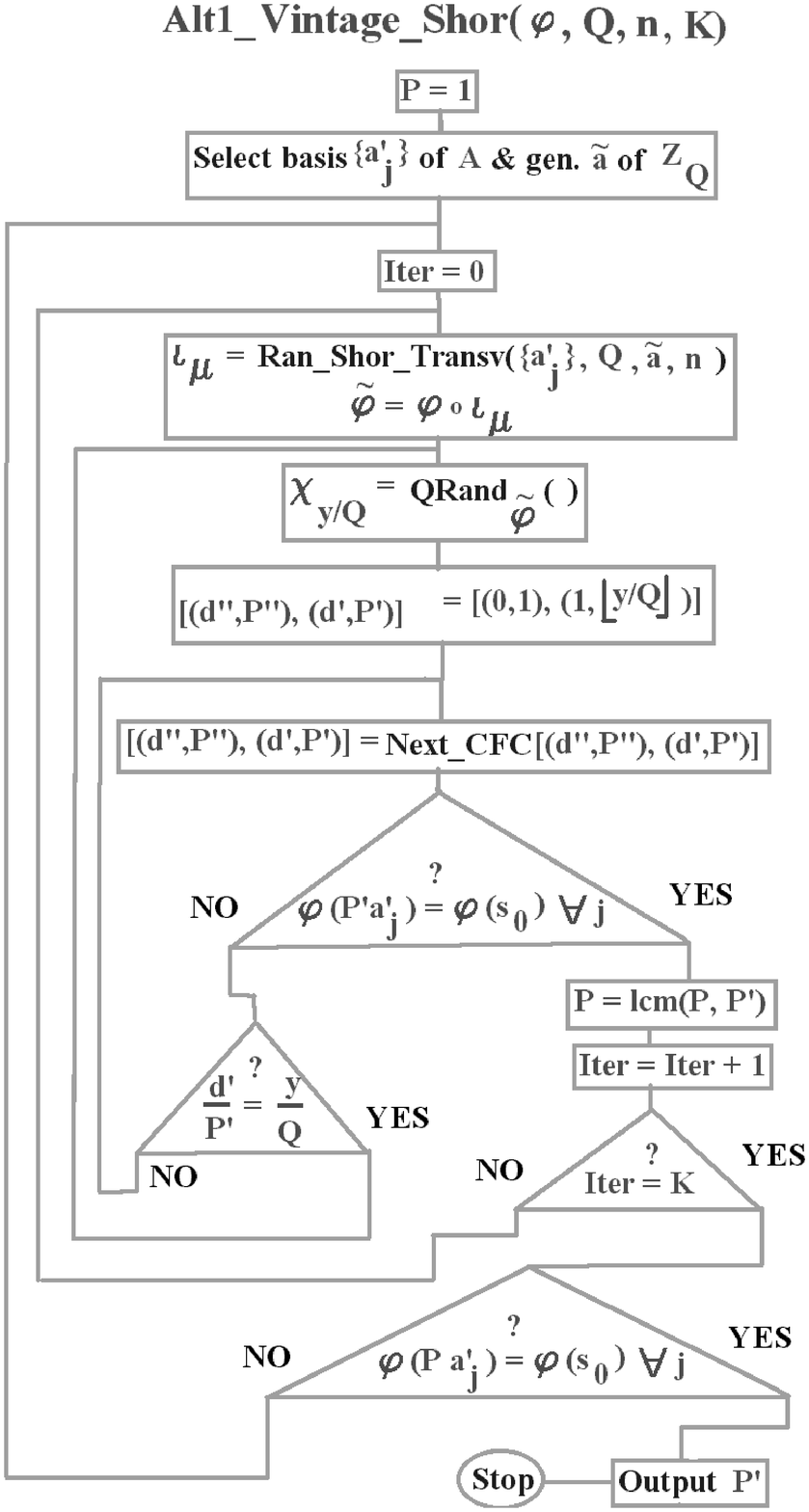}%
\\
\textbf{Figure 4. Flowchart for the second wandering Shor algorithm (a.k.a., a
vintage Shor algorithm). \ This algorithm finds the order }$P$\textbf{\ of a
maximal cyclic subgroup of the hidden quotient group }$H_{\varphi}$\textbf{.}%
\end{center}

\bigskip%
\begin{center}
\includegraphics[
height=7.075in,
width=4.4304in
]%
{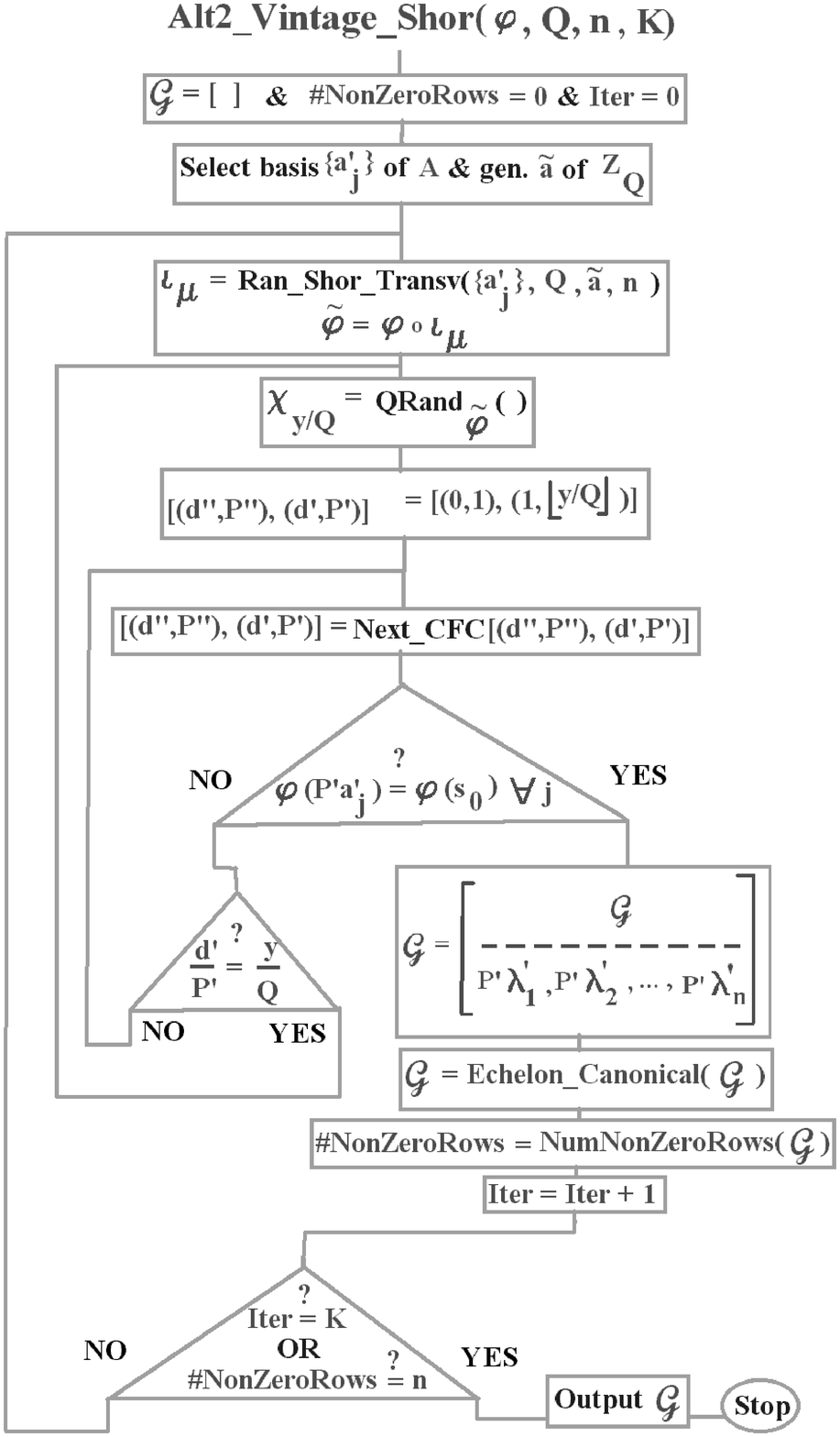}%
\\
\textbf{Figure 5. Flowchart for the third wandering Shor algorithm, a.k.a., a
vintage Shor algorithm. This algorithm finds the entire hidden quotient group
\ }$H_{\varphi}$\textbf{.}%
\end{center}

\bigskip

The algorithmic complexities of the above wandering Shor algorithms is given
in \cite{Lomonaco4}. \ For example, the first wandering Shor algorithm is of
time complexity%
\[
O\left(  n^{2}\left(  \operatorname{lg}N\right)  ^{3}\left(  \operatorname{lg}%
\operatorname{lg}N\right)  ^{n+1}\right)  \text{\ ,}%
\]
where $n$ is the rank of the free abelian group $A$. This can be readily
deduced from the abbreviated flowchart given in figure 6.

\bigskip%
\begin{center}
\includegraphics[
height=4.6345in,
width=4.1442in
]%
{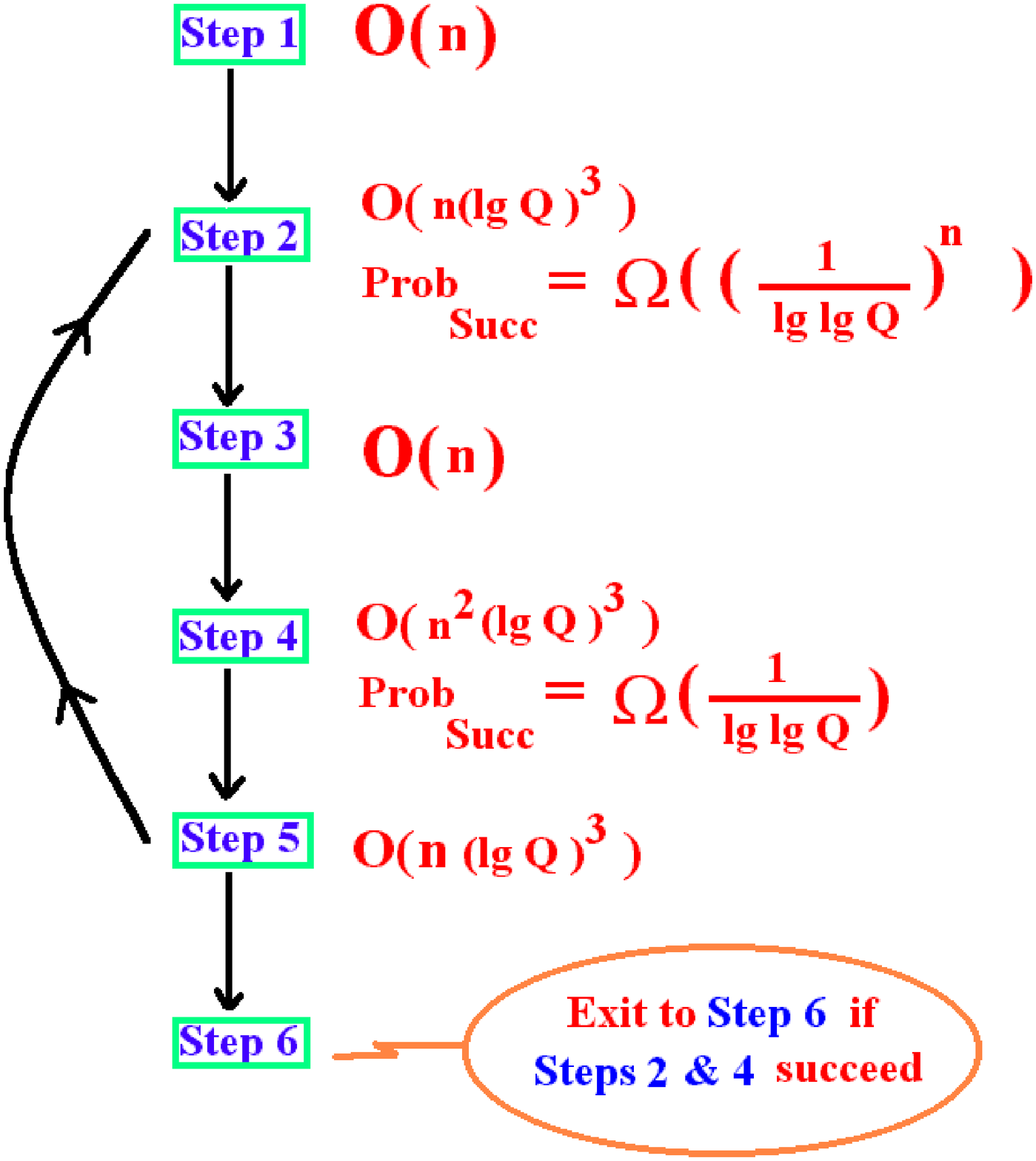}%
\\
\textbf{Figure 6. Abbreviated flowchart for the first wandering Shor
algorithm.}%
\end{center}

\bigskip

\section{Continuous (variable) Shor algorithms}

\bigskip

In in \cite{Lomonaco7} and in \cite{Lomonaco9}, the algorithmic primitives
found in above sections of this paper were used to create a class of
algorithms called continuous Shor algorithms. \ By a \textbf{continuous
variable Shor algorithm}, we mean a quantum hidden subgroup algorithm that
finds the hidden period $P$ of an admissible function $\varphi:\mathbb{R}%
\longrightarrow\mathbb{R}$ from the reals $\mathbb{R}$ to itself.

\bigskip

\begin{remark}
By an admissible function, we mean a function belonging to any sufficiently
well behaved class of functions. \ For example, the class of functions which
are Lebesgue integrable on every closed interval of $\mathbb{R}$. \ There, are
many other classes of functions that work equally as well.
\end{remark}

\bigskip

Actually, the papers \cite{Lomonaco7}, \cite{Lomonaco9} give in succession
three such continuous Shor algorithms, each successively more general than the previous.

\bigskip

For the first algorithm, we assume that the unknown hidden period $P$ is an
integer. \ The algorithm is then constructed by using rigged Hilbert
spaces\cite{Bohm1}, \cite{Gadella1}, linear combinations of Dirac delta
functions, and a subtle extension of the Fourier transform found in the
generic QHS subroutine \textsc{QRand}$(\varphi)$, which has been described
previously in section 4 of this paper. \ In Step 5 of \textsc{QRand}%
$(\varphi)$, the observable%
\[
A=%
{\displaystyle\int\limits_{-\infty}^{\infty}}
dy\dfrac{\left\lfloor Qy\right\rfloor }{Q}\left\vert y\right\rangle
\left\langle y\right\vert
\]
is measured, where $Q$ is an integer chosen so that $Q\geq2P^{2}$. \ It then
follows that the output of this algorithm is a rational $m/Q$ which is a
convergent of the continued fraction expansion of a rational of the form $n/P$.

\bigskip

The above quantum algorithm is then extended to a second quantum algorithm
that finds the hidden period $P$ of functions $\varphi:\mathbb{R}%
\longrightarrow\mathbb{R}$, where the unknown period $P$ is a rational.

\bigskip

Finally, the second algorithm is extended to a third algorithm which finds the
hidden period $P$ of functions $\varphi:\mathbb{R}\longrightarrow\mathbb{R}$,
when $P$ is an \textit{arbitrary real number}. We point out that for the third
and last algorithm to work, we must impose a very restrictive condition on the
map $\varphi:\mathbb{R}\longrightarrow\mathbb{R}$, i.e., the condition that
the map $\varphi$ is continuous.

\bigskip

\section{The quantum circle and the dual Shor algorithms.}

\bigskip

We have shown in previous sections how the mathematical primitives of pushing
and lifting can be used to create new quantum algorithms. \ In particular, we
have described how pushing and lifting can be used to derive new HSPs from an
HSP $\varphi:G\longrightarrow S$ on an arbitrary group $G$. \ We now see how
group duality can be exploited by these two primitives to create even more
quantum algorithms.

\bigskip%

\begin{center}
\includegraphics[
natheight=7.499600in,
natwidth=9.999800in,
height=2.2779in,
width=3.0268in
]%
{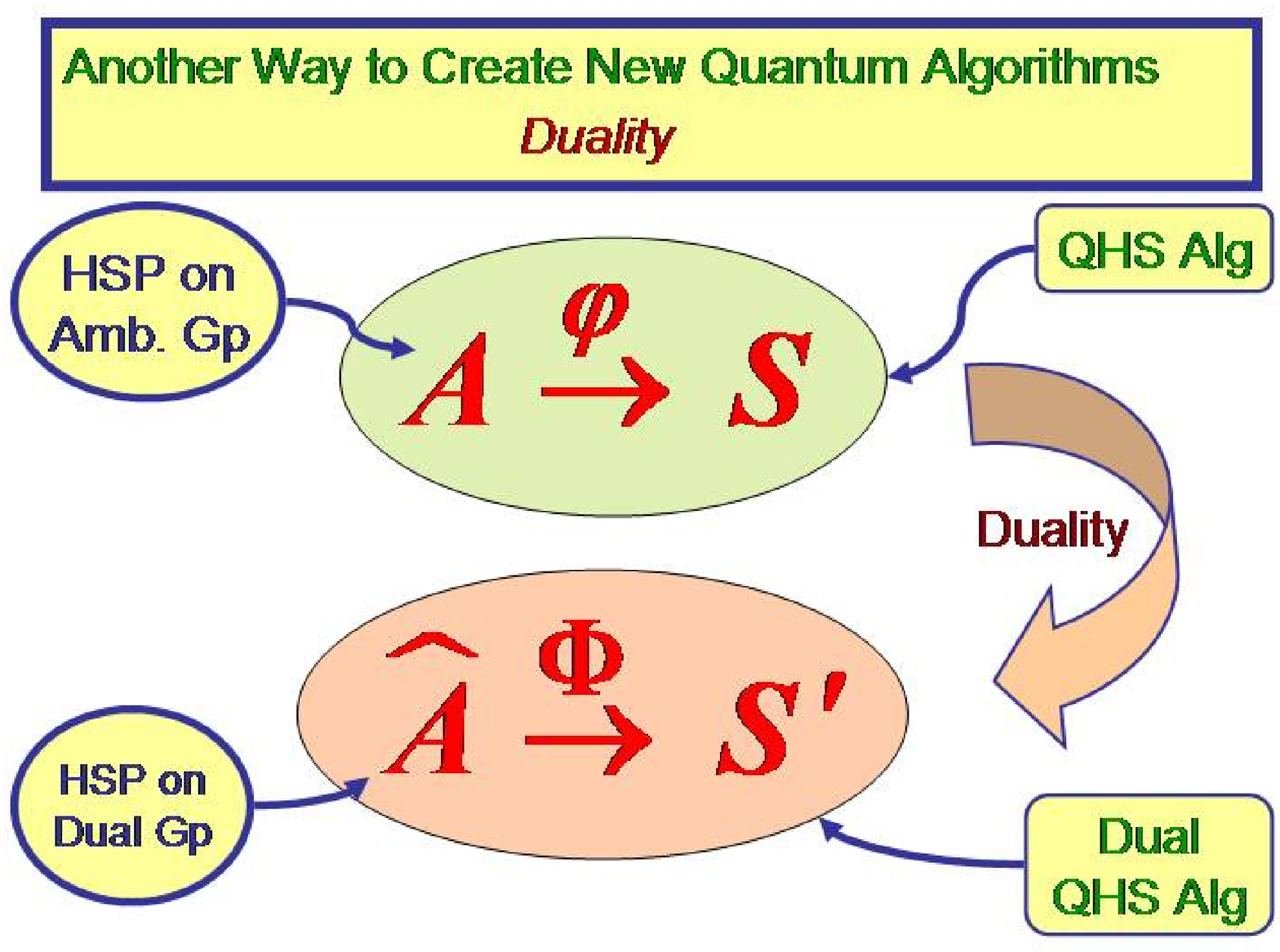}%
\\
\textbf{Figure 7. Using duality to create new QHS algorithms.}%
\end{center}

\bigskip

To this end, we assume that $G$ is an \textit{abelian} group. \ Hence, its
dual group of characters $\widehat{G}$ exists\footnote{If $G$ is non-abelian,
then its dual is not a group, but instead the representation algebra
$\mathcal{A}$ over the group ring $\mathbb{C}G$. \ The methods described in
this section can also be used to create new quantum algorithms for HSPs
$\Phi:\mathcal{A}\longrightarrow S$ on on the representation algebra
$\mathcal{A}$.}. \ It now follows that pushing and lifting can also be used to
derive new HSPs from an arbitrary HSP $\Phi:\widehat{G}\longrightarrow
S^{\prime}$ on the dual group $\widehat{G}$. \ In \cite{Lomonaco7}, this
method is used to create a number of new quantum algorithms derived from
Shor-like HSPs $\varphi:\mathbb{Z}\longrightarrow S$. \ 

\bigskip

A roadmap is shown in figure 8 of the developmental steps taken to find and to
create a new QHS algorithm on $\mathbb{Z}_{Q}$, which is (in the sense
described below) dual to Shor's original algorithm. \ We call the algorithm
developed in the final step of figure 8 the \textbf{dual Shor algorithm}.

\bigskip%
\begin{center}
\includegraphics[
natheight=7.499600in,
natwidth=9.999800in,
height=2.6524in,
width=3.5276in
]%
{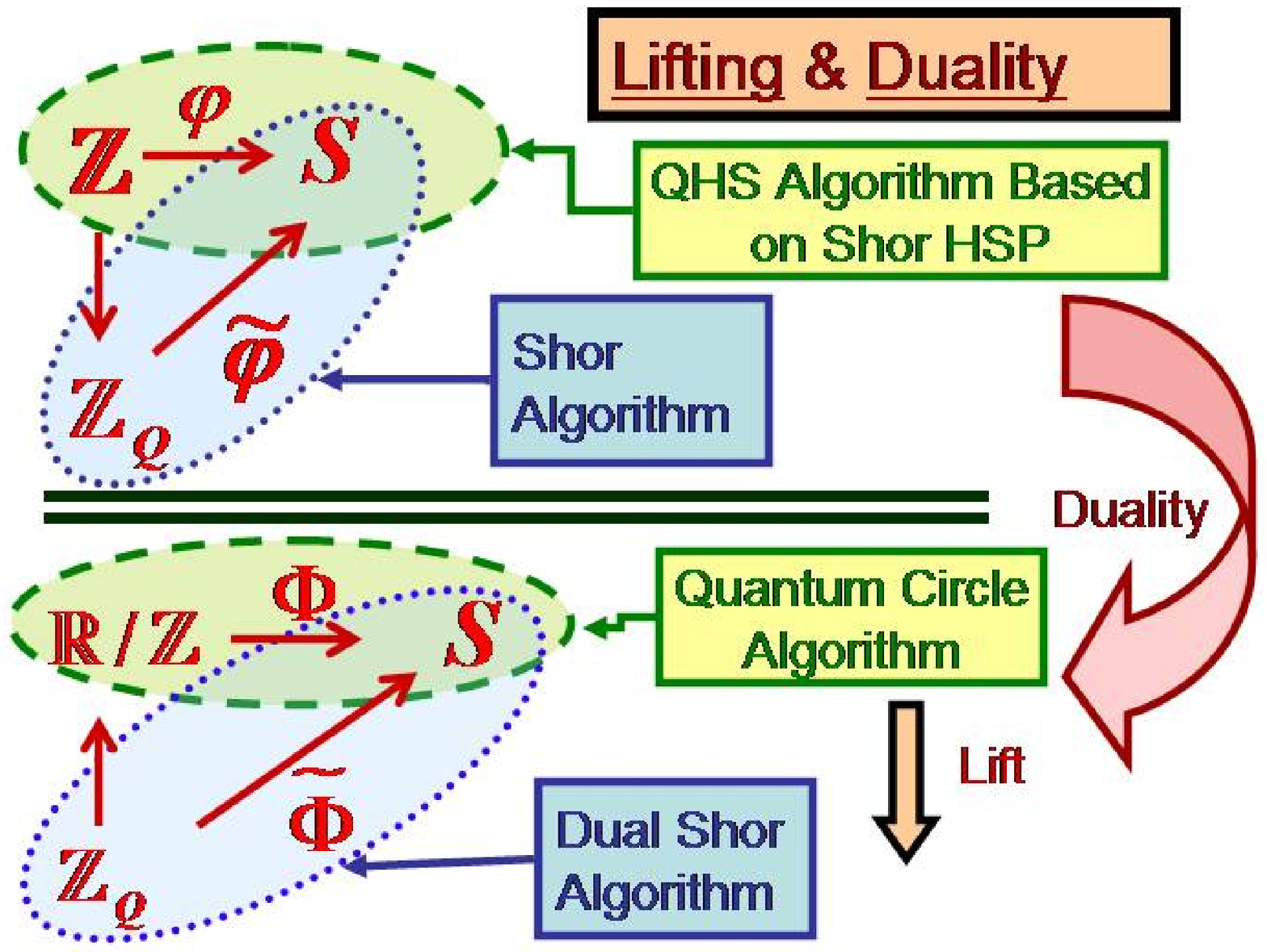}%
\\
\textbf{Figure 8. Roadmap for creating the dual Shor algorithm.}%
\end{center}

\bigskip

As indicated in figure 5, our first step is to create an intermediate QHS
algorithm based on a Shor-like HSP $\varphi:\mathbb{Z}\longrightarrow S$ from
the additive group of integers $\mathbb{Z}$ to a target set $S$. \ The
resulting algorithm "lives" in the infinite dimensional space $\mathcal{H}%
_{\mathbb{Z}}$ defined by the orthonormal basis $\left\{  \left\langle
n\right\vert :n\in\mathbb{Z}\right\}  $. \ This is a physically
unemplementable quantum algorithm created as a first steping stone in our
algorithmic development sequence. \ Intuitively, this algorithm can be viewed
as a "distillation" or a "purification" of Shor's original algorithm.

\bigskip

As a next step, \textbf{duality} is used to create the \textbf{quantum circle
algorithm}. \ This is accomplished by devising a QHS algorithm for an HSP
$\Phi:\mathbb{R}/\mathbb{Z}\longrightarrow S$ on the dual group $\mathbb{R}%
/\mathbb{Z}$ of the additive group of integers $\mathbb{Z}$. \ (By
$\mathbb{R}/\mathbb{Z}$, we mean the \textbf{additive group of reals
}$\mathbf{mod1}$, which is isomorphic to the multiplicative group $\left\{
e^{2\pi i\theta}:0\leq\theta<1\right\}  $, i.e., the \textbf{unit circle} in
the complex plane.) \ Once again, this is probably a physcally unemplementable
quantum algorithm\footnote{There is a possibility that the quantum circle
algorithm may have a physical implementation in terms of quantum optics.}.
\ But its utility lies in the fact that it leads to the physically
implementable quantum algorithm created in the last and final developmental
step, as indicated in figure 8. \ For in the final step, a physically
implementable QHS algorithm is created by \textbf{lifting} the HSP
$\Phi:\mathbb{R}/\mathbb{Z}\longrightarrow S$ to an HSP $\widetilde{\Phi
}:\mathbb{Z}_{Q}\longrightarrow S$. \ For the obvious reason, we call the
resulting algorithm a \textbf{dual Shor algorithm}.

\bigskip

For detailed descriptions of each of these quantum algorithms, i.e., the
"distilled" Shor, the quantum circle, and the dual Shor algorithms, the reader
is referred to \cite{Lomonaco7} and \cite{Lomonaco9}.

\bigskip

We give below brief descriptions of the quantum circle and the dual Shor algorithms.

\bigskip

For the \textbf{quantum circle algorithm}, we make use of the following spaces
(each of which is used in quantum optics):

\begin{itemize}
\item The rigged Hilbert space $H_{\mathbb{R}\mathbf{/}\mathbb{Z}}$ with
orthonormal basis $\left\{  \left\vert x\right\rangle :x\in\mathbb{R}%
/\mathbb{Z}\right\}  $. \ By \textquotedblleft orthonormal\textquotedblright%
\ we mean that $\left\langle x|y\right\rangle =\delta\left(  x-y\right)  $,
where \textquotedblleft\ $\delta$\textquotedblright\ denotes the Dirac delta
function. \ The elements of $H_{\mathbb{R}\mathbf{/}\mathbb{Z}}$ are
\textbf{formal integrals} of the form $\oint dx\ f(x)\left\vert x\right\rangle
$. \ (The physicist Dirac in his classic book\cite{Dirac1} on quantum
mechanics refers to these integrals as infinite sums. \ See also \cite{Bohm1}
and \cite{Gadella1}.)

\item The complex vector space $H_{\mathbb{Z}}$ of formal sums%
\[
\left\{  \sum\limits_{n=-\infty}^{\infty}a_{n}\left\vert \mathbf{n}%
\right\rangle :a_{n}\in C\;\;\forall n\in\mathbb{Z}\right\}
\]
with orthonormal basis $\left\{  \left\vert n\right\rangle :n\in Z\right\}  $.
\ By \textquotedblleft orthonormal\textquotedblright\ we mean that
$\left\langle n|m\right\rangle =\delta_{nm}$, where $\delta_{nm}$ denotes the
Kronecker delta.
\end{itemize}

\bigskip

We can now design an algorithm which solves the following hidden subgroup problem:

\bigskip

\noindent\textbf{Hidden Subgroup Problem for the Circle.} \ \textit{Let }%
$\Phi:\mathbb{R}/\mathbb{Z}\longrightarrow\mathbb{C}$\textit{ be an admissible
function from the circle group }$\mathbb{R}/\mathbb{Z}$\textit{ to the complex
numbers }$\mathbb{C}$\textit{ with hidden rational period }$\alpha
\in\mathbb{Q}/\mathbb{Z}$\textit{, where }$\alpha\in\mathbb{Q}/\mathbb{Z}%
$\textit{ denotes the rational circle, i.e., the rationals }%
$\operatorname*{mod}1$\textit{.}

\bigskip

\begin{remark}
By an admissible function, we mean a function belonging to any sufficiently
well behaved class of functions. \ For example, the class of functions which
are Lebesgue integrable on $\mathbb{R}/\mathbb{Z}$. \ There, are many other
classes of functions that work equally as well.
\end{remark}

\bigskip

\begin{proposition}
If $\alpha=a_{\mathbf{1}}/a_{\mathbf{2}}$ (with $\gcd\left(  a_{1}%
,a_{2}\right)  =1$) is a rational period of a function $\Phi:\mathbb{R}%
/\mathbb{Z}\longrightarrow C$, then $1/a_{\mathbf{2}}$is also a period of
$\Phi$. \ Hence, the minimal rational period of $\Phi$ is always a reciprocal
integer $\operatorname*{mod}1$.
\end{proposition}

\bigskip

The following quantum algorithm finds the reciprocal integer period of the
function $\Phi$.

\bigskip

\begin{center}
\textsc{Circle-Algorithm}$(\Phi)$\bigskip
\end{center}

\begin{itemize}
\item[\textbf{Step 0.}] Initialization%
\[
\left\vert \psi_{0}\right\rangle =\left\vert 0\right\rangle \left\vert
0\right\rangle \in H_{\mathbb{Z}}\otimes H_{\mathbb{C}}%
\]
\bigskip

\item[\textbf{Step 1.}] Application of the inverse Fourier transform
$\mathcal{F}^{-1}\otimes1$%
\[
\left\vert \psi_{1}\right\rangle =%
{\displaystyle\int}
dx\ e^{\mathbf{2}\pi i\cdot0}\left\vert x\right\rangle \left\vert
0\right\rangle =%
{\displaystyle\int}
dx\ \left\vert x\right\rangle \left\vert 0\right\rangle \in H_{\mathbb{R}%
\mathbf{/}\mathbb{Z}}\otimes H_{\mathbb{C}}%
\]
\bigskip

\item[\textbf{Step 2.}] Step 2. Application of the unitary transformation
$U_{\varphi}:\left\vert x\right\rangle \left\vert u\right\rangle
\mapsto\left\vert x\right\rangle \left\vert u+\Phi(x)\right\rangle $%
\[
\left\vert \psi_{2}\right\rangle =%
{\displaystyle\int}
dx\ \left\vert x\right\rangle \left\vert \Phi(x)\right\rangle \in
H_{\mathbb{R}\mathbf{/}\mathbb{Z}}\oplus H_{\mathbb{C}}%
\]
\bigskip

\item[\textbf{Step 3.}] Application of the Fourier transform $\mathcal{F}%
\otimes1$
\end{itemize}

\bigskip

\begin{remark}
Remark. Letting $x_{m}=x-\dfrac{m}{a}$, we have%
\begin{gather*}%
{\displaystyle\int}
dx\ e^{\mathbf{2}\pi inx}\left\vert \Phi(x)\right\rangle =\sum\limits_{m=0}%
^{a-1}%
{\displaystyle\int\limits_{m/a}^{\mathbf{(}m+1)/a}}
dx\ e^{-2\pi inx}\left\vert \Phi(x)\right\rangle \\
\qquad\qquad\qquad\;\;=\sum\limits_{m=0}^{a-1}%
{\displaystyle\int\limits_{0}^{\mathbf{1}/a}}
dx_{m}\ e^{-2\pi in\left(  x_{m}+\dfrac{m}{a}\right)  }\left\vert \Phi\left(
x_{m}+\dfrac{m}{a}\right)  \right\rangle \\
\qquad\qquad\qquad\;\;=\left(  \sum\limits_{m=0}^{a-1}e^{-2\pi inm/a}\right)
{\displaystyle\int\limits_{0}^{\mathbf{1}/a}}
dx\ e^{-2\pi inx}\left\vert \Phi\left(  x\right)  \right\rangle
\end{gather*}
where $1/a$ is the unknown reciprocal period. But%
\[%
{\displaystyle\sum\limits_{m=0}^{a-1}}
e^{-2\pi inm/a}=a\delta_{n=0\operatorname*{mod}a}=\left\{
\begin{array}
[c]{ll}%
a & \text{if \ }n=0\operatorname*{mod}a\\
0 & \text{otherwise}%
\end{array}
\right.
\]
Hence,%
\begin{gather*}
\left\vert \psi_{\mathbf{3}}\right\rangle =\sum\limits_{n\in\mathbb{Z}%
}\left\vert n\right\rangle
{\displaystyle\int}
dx\ e^{-2\pi inx}\left\vert \Phi\left(  x\right)  \right\rangle =\left(
\sum\limits_{n\in Z}\left\vert n\right\rangle \delta_{n=0\operatorname*{mod}%
a}\right)
{\displaystyle\int\limits_{0}^{\mathbf{1}/a}}
dx\ e^{-2\pi inx}\left\vert \Phi\left(  x\right)  \right\rangle \\
\quad\;\,=\left(  \sum\limits_{\ell\in Z}\left\vert \ell a\right\rangle
\right)  \left(
{\displaystyle\int\limits_{0}^{\mathbf{1}/a}}
dx\ e^{-2\pi inx}\left\vert \Phi\left(  x\right)  \right\rangle \right)
=\sum\limits_{\ell\mathbf{\in}\mathbb{Z}}\left\vert \ell a\right\rangle
\left\vert \Omega\left(  \ell a\right)  \right\rangle
\end{gather*}
\bigskip
\end{remark}

\begin{itemize}
\item[\textbf{Step 4.}] Measurement of%
\[
\left\vert \psi_{\mathbf{3}}\right\rangle \mathbb{=}\sum\limits_{\ell
\in\mathbb{Z}}\left\vert \ell a\right\rangle \left\vert \mathbf{\Omega}\left(
\ell\mathbf{a}\right)  \right\rangle \mathbb{\in H}_{\mathbb{Z}}%
\mathbb{\otimes H}_{\mathbb{C}}%
\]
with respect to the observable%
\[
\sum\limits_{n\in\mathbb{Z}}n\left\vert n\right\rangle \left\langle
n\right\vert
\]
to produce a random eigenvalue $\ell a$.
\end{itemize}

\bigskip

\begin{remark}
The above quantum circle algorithm can be extended to a quantum algorithm
which finds the hidden period $\alpha$ of a function $\Phi:\mathbb{R}%
/\mathbb{Z}\longrightarrow\mathbb{C}$, when $\alpha$ is an arbitrary real
number $\operatorname*{mod}1$. \ But in creating this extended quantum
algorithm, a very restrictive condition must be imposed on the map
$\Phi:\mathbb{R}/\mathbb{Z}\longrightarrow\mathbb{C}$, namely, the condition
that $\Phi$ be continuous.
\end{remark}

\bigskip

We now give a brief description of the \textbf{dual Shor algorithm}.

\bigskip

The dual Shor algorithm is a QHS algorithm created by making a discrete
approximation of the quantum circle algorithm. \ More specifically, it is
created by lifting the QHS circle algorithm for $\varphi:\mathbb{R}%
/\mathbb{Z}\longrightarrow\mathbb{C}$ to the finite cyclic group
$\mathbb{Z}_{Q}$, as illustrated in the commutative diagram given below:%
\[%
\begin{array}
[c]{ll}%
\;\,\mathbb{Z}_{Q} & \\
\mu\downarrow\;\; & \;\;\searrow\widetilde{\varphi}=Push\left(  \varphi
\right)  =\varphi\circ\mu\\
\mathbb{R}\mathbf{/}\mathbb{Z} & \longrightarrow\;\;S
\end{array}
\]
Intuitively, just as in Shor's algorithm,\ the circle group $\mathbb{R}%
/\mathbb{Z}$ is "approximated" with the finite cyclic group $\mathbb{Z}_{Q}$,
where the group $\mathbb{Z}_{Q}$ is identified with the additive group%
\[
\left\{  \dfrac{0}{Q},\dfrac{1}{Q},\ldots.,\dfrac{Q-1}{Q}\right\}
\operatorname*{mod}1\text{ \ ,}%
\]
and where the hidden subgroup $\mathbb{Z}_{P}$ is identified with the additive
group%
\[
\left\{  \dfrac{0}{P}\mathbf{,}\dfrac{1}{P}\mathbf{,}\ldots\mathbf{.}%
\mathbf{,}\dfrac{P-1}{P}\right\}  \operatorname*{mod}1\text{ \ ,}%
\]
with $P=a_{2}$.

\bigskip

This is a physically implementable quantum algorithm. \ In a certain sense, it
is actually faster than Shor's algorithm. \ For the last step of Shor's
algorithm uses the standard continued fraction algorithm to determine the
unknown period. \ On the other hand, the last step of the dual Shor algorithm
uses the much faster Euclidean algorithm to compute the greatest common
divisor of the integers $\ell_{1}a,\ell_{2}a,\ell_{3}a,\ldots$, thereby
determining the desired reciprocal integer period $1/a$. \ For more details,
please refer to \cite{Lomonaco7} and \cite{Lomonaco9}.

\bigskip

\section{A QHS algorithm for Feynman integrals.}

We now discuss a QHS algorithm based on Feynman path integrals. \ This quantum
algorithm was developed at the Mathematical Sciences Research Institute (MSRI)
in Berkeley, California when one of the authors of this paper was challenged
with an invitation to give a talk on the relation between Feynmann path
integrals and quantum computing at an MSRI conference on Feynman path integrals.

\bigskip

Until recently, both authors of this paper thought that the quantum algorithm
to be described below was a highly speculative quantum algorithm. \ For the
existence of Feynman path integrals is very difficult (if not impossible) to
determine in a mathematically rigorous fashion. \ But surprisingly, Jeremy
Becnel in his doctoral dissertation \cite{Becnel1} actually succeeded in
creating a firm mathematical foundation for this algorithm. \ 

\bigskip

We should mention, however, that the physical implementability of this
algorithm is still yet to be determined.

\bigskip

\begin{definition}
Definition. Let \textsc{Paths} be the real vector space of all continuous
paths $x:\left[  0,1\right]  \longrightarrow\mathbb{R}^{n}$ which are $L^{2}$
with respect to the inner product%
\[
x\cdot y=%
{\displaystyle\int_{0}^{1}}
ds\ x(s)y(s)
\]
with scalar multiplication and vector sum defined as

\begin{itemize}
\item $\left(  \lambda x\right)  \left(  s\right)  =\lambda x\left(  s\right)
$

\item $\left(  x+y\right)  \left(  s\right)  =x\left(  s\right)  +y\left(
s\right)  $
\end{itemize}
\end{definition}

\bigskip

We wish to create a QHS algorithm for the following hidden subgroup problem:

\bigskip

\noindent\textbf{Hidden Subgroup Problem for }\textsc{Paths}. \ \textit{Let
}$\varphi:\ $\textsc{Paths}$\ \longrightarrow C$\textit{ be a functional with
a hidden subspace }$V$\textit{ of }\textsc{Paths}\textit{ such that}%
\[
\varphi\left(  x+v\right)  =\varphi\left(  x\right)  \;\;\forall v\in V
\]

\bigskip

Our objective is to create a QHS algorithm which solves the above problem,
i.e., which finds the hidden subspace $V$.

\bigskip

\begin{definition}
Let $H_{\text{\textsc{Paths}}}$ be the rigged Hilbert space with orthonormal
basis $\left\{  \left\vert x\right\rangle :x\in\text{\textsc{Paths}}\right\}
$, and with bracket product $\left\langle x|y\right\rangle =\delta\left(
x-y\right)  $.
\end{definition}

\bigskip

We will use the following observation to create the QHS algorithm:

\bigskip

\noindent\textbf{Observation.} \textsc{Paths}$\ =%
{\displaystyle\bigcup\nolimits_{v\in V}}
\left(  v+V^{\mathbf{\perp}}\right)  $\textit{, where }$V^{\mathbf{\perp}}%
$\textit{ denotes the orthogonal complement of the hidden vector subspace }%
$V$\textit{.}

\bigskip

The QHS algorithm for Feynman path integral is given below:

\bigskip

\begin{center}
\textsc{Feynman}$\left(  \varphi\right)  $

\end{center}

\begin{itemize}
\item[\textbf{Step 0.}] Initialize%
\[
\left\vert \psi_{0}\right\rangle =\left\vert 0\right\rangle \left\vert
0\right\rangle \in H_{Paths}\otimes H_{\mathbb{C}}%
\]
\bigskip

\item[\textbf{Step 1.}] Apply $\mathcal{F}^{-1}\otimes1$%
\[
\left\vert \psi_{1}\right\rangle =%
{\displaystyle\int\limits_{\text{\textsc{Paths}}}}
\mathcal{D}x\;e^{2\pi ix\cdot0}\left\vert x\right\rangle \left\vert
0\right\rangle =%
{\displaystyle\int\limits_{\text{\textsc{Paths}}}}
\mathcal{D}x\ \left\vert x\right\rangle \left\vert 0\right\rangle
\]
\bigskip

\item[\textbf{Step 2.}] Apply $U_{\mathbf{\varphi}}:\left\vert x\right\rangle
\left\vert u\right\rangle \mapsto\left\vert x\right\rangle \left\vert
u+\varphi(x)\right\rangle $%
\[
\left\vert \psi_{\mathbf{2}}\right\rangle =%
{\displaystyle\int\limits_{\text{\textsc{Paths}}}}
\mathcal{D}x\ \left\vert x\right\rangle \left\vert \varphi(x)\right\rangle
\]
\bigskip

\item[\textbf{Step 3.}] Apply $\mathcal{F}\otimes1$%
\[%
\begin{array}
[c]{lll}%
\left\vert \psi_{3}\right\rangle  & = &
{\displaystyle\int\limits_{\text{\textsc{Paths}}}}
\mathcal{D}y%
{\displaystyle\int\limits_{\text{\textsc{Paths}}}}
\mathcal{D}x\ e^{-2\pi ix\cdot y}\left\vert y\right\rangle \left\vert
\varphi\left(  x\right)  \right\rangle \\
& = &
{\displaystyle\int\limits_{\text{\textsc{Paths}}}}
\mathcal{D}y\ \left\vert y\right\rangle
{\displaystyle\int\limits_{\text{\textsc{Paths}}}}
\mathcal{D}x\ e^{-2\pi ix\cdot y}\left\vert \varphi\left(  x\right)
\right\rangle
\end{array}
\]
But
\begin{gather*}%
{\displaystyle\int\limits_{\text{\textsc{Paths}}}}
\mathcal{D}x\ e^{-2\pi ix\cdot y}\left\vert \varphi\left(  x\right)
\right\rangle =%
{\displaystyle\int\limits_{V}}
\mathcal{D}v%
{\displaystyle\int\limits_{v+V^{\mathbf{\perp}}}}
\mathcal{D}x\ e^{-2\pi ix\cdot y}\left\vert \varphi\left(  x\right)
\right\rangle \\
\qquad\qquad\qquad\qquad\;\;=%
{\displaystyle\int\limits_{V}}
\mathcal{D}v%
{\displaystyle\int\limits_{V^{\mathbf{\perp}}}}
\mathcal{D}x\ e^{-2\pi i\left(  \mathbf{v}\mathbf{+}\mathbf{x}\right)  \cdot
y}\left\vert \varphi\left(  v+x\right)  \right\rangle \\
\qquad\qquad\qquad\qquad\;\;=%
{\displaystyle\int\limits_{V}}
\mathcal{D}v\ e^{-2\pi iv\cdot y}%
{\displaystyle\int\limits_{V^{\mathbf{\perp}}}}
\mathcal{D}x\ e^{-2\pi ix\cdot y}\left\vert \varphi\left(  x\right)
\right\rangle
\end{gather*}
However,%
\[%
{\displaystyle\int\limits_{V}}
\mathcal{D}v\ e^{-2\pi iv\cdot y}=%
{\displaystyle\int\limits_{V^{\mathbf{\perp}}}}
\mathcal{D}u\ \delta\left(  y-u\right)
\]
So,
\begin{gather*}
\left\vert \psi_{3}\right\rangle =%
{\displaystyle\int\limits_{\text{\textsc{Paths}}_{n}}}
\mathcal{D}y\ \left\vert y\right\rangle
{\displaystyle\int\limits_{V}}
\mathcal{D}v\ e^{-2\pi iv\cdot y}%
{\displaystyle\int\limits_{V^{\mathbf{\perp}}}}
\mathcal{D}x\ e^{-2\pi ix\cdot y}\left\vert \varphi\left(  x\right)
\right\rangle \\
\quad\;\;=%
{\displaystyle\int\limits_{\text{\textsc{Paths}}_{n}}}
\mathcal{D}y\ \left\vert y\right\rangle
{\displaystyle\int\limits_{V^{\mathbf{\perp}}}}
\mathcal{D}u\ \delta\left(  y-u\right)
{\displaystyle\int\limits_{V^{\mathbf{\perp}}}}
\mathcal{D}x\ e^{-2\pi ix\cdot y}\left\vert \varphi\left(  x\right)
\right\rangle \\
\quad\;\;=%
{\displaystyle\int\limits_{V^{\mathbf{\perp}}}}
\mathcal{D}u\ \left\vert u\right\rangle
{\displaystyle\int\limits_{V^{\mathbf{\perp}}}}
\mathcal{D}x\ e^{-2\pi ix\cdot u}\left\vert \varphi\left(  x\right)
\right\rangle \\
\quad\;\;=%
{\displaystyle\int\limits_{V^{\mathbf{\perp}}}}
\mathcal{D}u\ \left\vert u\right\rangle \left\vert \Omega\left(  u\right)
\right\rangle
\end{gather*}
\bigskip

\item[\textbf{Step 4.}] Measure%
\[
\left\vert \psi_{3}\right\rangle =%
{\displaystyle\int\limits_{V^{\mathbf{\perp}}}}
\mathcal{D}u\ \left\vert u\right\rangle \left\vert \Omega\left(  u\right)
\right\rangle
\]
with respect to the observable%
\[
A=%
{\displaystyle\int\limits_{\text{\textsc{Paths}}}}
\mathcal{D}w\ \left\vert w\right\rangle \left\langle w\right\vert
\]
to produce a random element of $V^{\mathbf{\perp}}$
\end{itemize}

\bigskip

The above algorithm suggests an intriguing question. Can the above QHS Feynman
integral algorithm be modified in such a way as to create a quantum algorithm
for the Jones polynomial? \ In other words, can it be modified by replacing
\textsc{Paths} with the space of gauge connections, and making suitable modifications?

\bigskip

This question is motivated by the fact that the integral over gauge
transformations%
\[
\widehat{\psi}\left(  K\right)  =%
{\displaystyle\int}
\mathcal{D}A\ \psi\left(  A\right)  \mathcal{W}_{K}\left(  A\right)
\]
looks very much like a Fourier transform, where%
\[
\mathcal{W}_{K}\left(  A\right)  =tr\left(  P\exp\left(  \oint_{K}A\right)
\right)
\]
denotes the \textbf{Wilson loop} over the knot $K$.

\bigskip

\section{QHS algorithms on free groups}

\bigskip

In this and the following section of this paper, our objective is to show that
a free group is the the most natural domain for QHS algorithms. \ In
retrospect, this is not so surprising if one takes a discerning look at Shor's
factoring algorithm. \ For in section 6, we have seen that Shor's algorithm is
essentially a QHS algorithm on the free group $\mathbb{Z}$ which has been
pushed onto the finite group $\mathbb{Z}_{Q}$.

\bigskip

In particular, let $\varphi:G\longrightarrow S$ be a map with hidden subgroup
structure from a finitely generated (f.g.) group $G$ to a set $S$. \ We assume
that the hidden subgroup $K$ is a normal subgroup of $G$ of finite index.
\ Then the objectives of this section are to demonstrate the following:

\bigskip

\begin{itemize}
\item Every hidden subgroup problem (HSP) $\varphi:G\longrightarrow S$ on an
arbitray f.g. group $G$ can be lifted to an HSP $\widetilde{\mathbf{\varphi}%
}:F\longrightarrow S$ on a free group $F$ of finite rank.

\item Moreover, a solution for the lifted HSP $\widetilde{\mathbf{\varphi}%
}:F\longrightarrow S$ is for all practical purposes the same as the solution
for the original HSP $\varphi:G\longrightarrow S$.
\end{itemize}

\bigskip

\textit{Thus, one need only investigate QHS algorithms for free groups of
finite rank!}

\bigskip

Before we can describe the above results, we need to review a number of
definitions. \ We begin with the definition of a free group:

\bigskip

\begin{definition}
[Universal Definition]A group $F$ is said to be \textbf{free} of finite rank
$n$ if there exists a finite set of $n$ generators $X=\left\{  x_{1}%
,x_{2},\ldots,x_{n}\right\}  $ such that, for every group $G$ and for every
map $f:X\longrightarrow G$ of the set $X$ into the group $G$, the map $f$
extends to a morphism $\widetilde{f}:F\longrightarrow G$. \ We call the set
$X$ a \textbf{free basis} of the group $F$, and frequently denote the group
$F$ by $F\left(  x_{1},x_{2},\ldots,x_{n}\right)  $, . It follows from this
definition that the morphism $\widetilde{f}$ is unique.
\end{definition}

\bigskip

The intuitive idea encapsulated by this definition is that a free group is an
unconstrained group (very much analogous to a physical system without boundary
conditions.) \ In other words, a group is free provided it has a set of
generators such that the only relations among those generators are those
required for $F$ to be a group. \ For example,

\begin{itemize}
\item $x_{i}x_{i}^{\mathbf{-}1}=1$ is an allowed relation

\item $x_{i}x_{j}=x_{j}x_{i}$ is not an allowed relation for $i\neq j$

\item $x_{i}^{3}=1$ is not an allowed relation
\end{itemize}

\bigskip

As an immediate consequence of the above definition, we have the following proposition:

\bigskip

\begin{proposition}
Let $G$ be an arbitrary f.g. group with finite set of $n$ generators $\left\{
g_{1},g_{2},\ldots,g_{n}\right\}  $, and let $F=F\left(  x_{1},x_{2}%
,\ldots,x_{n}\right)  $ be the free group of rank $n$ with free basis
$\left\{  x_{1},x_{2},\ldots,x_{n}\right\}  $. \ 

Then by the above definition, the map $x_{j}\longmapsto g_{j}$ \ $\left(
j=1,2,\ldots,n\right)  $ induces a uniques epimorphism $\nu:F\longrightarrow
G$ from $F$ onto $G$. \ With this epimorphism, every HSP $\varphi
:G\longrightarrow S$ on the group $G$ uniquely lifts to the HSP $\widetilde
{\mathbf{\varphi}}=\varphi\circ\nu:F\longrightarrow S$ on the free group $F$. \ 

Moreover, if $K$ and $\widetilde{K}$ are the hidden subgroups of the HSPs
$\varphi$ and $\widetilde{\mathbf{\varphi}}$, respectively, the corresponding
hidden quotient groups $G/K$ and $F/\widetilde{K}$ of these two HSPs are
isomorphic. \ Hence, every solution of the HSP $\widetilde{\mathbf{\varphi}%
}:F\longrightarrow S$ immediately produces a solution of the original HSP
$\varphi:G\longrightarrow S$.
\end{proposition}

\bigskip

We close this section with the defintion of a group resentation, a concept
that will be needed in the next section for generalizing Shor's algorithm to
free groups.

\bigskip

\begin{definition}
Let $G$ be a group. \ A \textbf{group presentation}%
\[
\left(  x_{1},x_{2},\ldots,x_{n}:r_{1},r_{2},\ldots,r_{m}\right)
\]
for $G$ is a set of free generators $x_{1},x_{2},\ldots,x_{n}$ of a free group
$F$ and a set of words $r_{1},r_{2},\ldots,r_{n}$ in $F\left(  x_{1}%
,x_{2},\ldots,x_{n}\right)  $, called \textbf{relators}, such that the group
$G$ is isomorphic to the quotient group $F\left(  x_{1},x_{2},\ldots
,x_{n}\right)  /Cons\left(  r_{\mathbf{1}},r_{2},\ldots,r_{n}\right)  $, where
$Cons\left(  r_{1},r_{2},\ldots,r_{n}\right)  $, called the
\textbf{consequence} of $r_{1},r_{2},\ldots,r_{n}$, is the smallest normal
subgroup of $F\left(  x_{1},x_{2},\ldots,x_{n}\right)  $ containing the
relators $r_{1},r_{2},\ldots,r_{n}$.
\end{definition}

\bigskip

The intuition captured by the above definition is that $x_{1},x_{2}%
,\ldots,x_{n}$ are the generators of $G$, and $r_{1}=1,r_{2}=1,\ldots,r_{n}=1$
is a complete set of relations among these generators, i.e., every relation
among the generators of $G$ is a \textbf{consequence} of (derivable from) the
relations $r_{1}=1,r_{2}=1,\ldots,r_{n}=1$. \ For example,

\begin{itemize}
\item $\left(  x_{1},x_{2},\ldots,x_{n}:\right)  $ and $\left(  x_{1}%
,x_{2},\ldots,x_{n}:x_{1}x_{1}^{-1},x_{2}^{5}x_{2}^{\mathbf{-}5},x_{3}%
x_{4}x_{4}^{\mathbf{-}1}x_{3}^{\mathbf{-}1}\right)  $ are both presentations
of the free group $F\left(  x_{1},x_{2},\ldots,x_{n}\right)  $

\item $\left(  x:x^{Q}\right)  $ and $\left(  x:x^{a},x^{b}\right)  $ are both
presentations of the cyclic group $\mathbb{Z}_{Q}$of order $Q$, where $a$ and
$b$ are integers such that $\gcd\left(  a,b\right)  =Q$.

\item $\left(  x_{1},x_{2}:x_{1}^{3},x_{2}^{2},\left(  x_{1}x_{2}\right)
^{2}\right)  $ is a presentation of the symmetric group $S_{3}$ on three symbols.
\end{itemize}

\bigskip

\section{Generalizing Shor's algorithm to free groups}

\bigskip

The objective of this section is to generalize Shor's algorithm to free groups
of finite rank\footnote{We remind the reader that, in section 6, we showed
that Shor's algorithm is essentially a QHS algorithm on the free group
$\mathbb{Z}$ of rank 1 constructed by a push onto the cyclic group
$\mathbb{Z}_{Q}$. \ In light of this and of the results outlined in the
previous section, it is a natural objective to generalize Shor's algorithm to
free groups of finite rank.}. \ The chief obstacle to accomplishing this goal
is finding a correct generalization of the Shor transversal
\[%
\begin{array}
[c]{cccc}%
\mathbb{Z}_{Q} & \overset{\tau}{\longrightarrow} & \mathbb{Z} & \\
n\operatorname{mod}Q & \longmapsto & n & \left(  \text{\ }0\leq n<Q\right)
\end{array}
\]
Unfortunately, there appear to be few mathematical clues indicating how to go
about making such a generalization. \ However, as we shall see, the
generalization of the Shor transversal to the transversal found in the
wandering Shor algorithm does provide a crucial clue, suggesting that a
generalized Shor transversal must be a 2-sided Schreier transversal. \ (See
section 7.)

\bigskip

We begin by formulating a constructive approach to free groups:

\bigskip

\begin{definition}
Let $F\left(  x_{1},x_{2},\ldots,x_{n}\right)  $ be a free group with free
basis $x_{1},x_{2},\ldots,x_{n}$. \ Then a \textbf{word} is a finite string of
the symbols $x_{1},x_{1}^{-1},x_{2},x_{2}^{-1},\ldots,x_{n},x_{n}^{-1}$. \ A
\textbf{reduced word} is a word in which there is no substring of the form
$x_{j}x_{j}^{-1}$ or $x_{j}^{-1}x_{j}$. \ Two words are said to be
\textbf{equivalent} if one can be transformed into the other by applying a
finite number of substring insertions or deletions of the form $x_{j}%
x_{j}^{-1}$ or $x_{j}^{-1}x_{j}$. \ We denote an \textbf{arbitrary word} $w$
by $w=a_{1}a_{2}\cdots a_{\ell}$ , where each $a_{j}=x_{k_{j}}^{\pm1}$. \ The
\textbf{length} $\left\vert w\right\vert $ of a word $w=a_{1}a_{2}\cdots
a_{\ell}$ is number of symbols $x_{k_{j}}^{\pm1}$ that appear in $w$, i.e.,
$\left\vert w\right\vert =\ell$.
\end{definition}

\bigskip

For example, $x_{2}x_{1}^{-1}x_{1}x_{1}^{\mathbf{-}1}x_{5}^{-1}x_{5}%
^{\mathbf{-}1}x_{5}^{\mathbf{-}1}x_{5}$ is a word of length $8$ which is
equivalent to the reduced word $x_{2}x_{1}^{\mathbf{-}1}x_{5}^{\mathbf{-}%
1}x_{5}^{\mathbf{-}1}$ of length $4$.

\bigskip

It easily follows that:

\bigskip

\begin{proposition}
A free group $F\left(  x_{1},x_{2},\ldots,x_{n}\right)  $ is simply the set of
reduced words together with the obvious definition of product, i.e.,
concatenation followed by full reduction.
\end{proposition}

\bigskip

We can now use this constructive approach to create a special kind of
transversal $\tau:G\longrightarrow F$ of an epimorphism $\nu:F\longrightarrow
G$, called a 2-sided Scheier transversal\cite{Hall1}:

\bigskip

\begin{definition}
A set $\mathcal{W}$ of reduced words in a free group $F=F\left(  x_{1}%
,x_{2},\ldots,x_{n}\right)  $ is said to be a \textbf{2-sided Schreier system} provided

\begin{itemize}
\item The empty word $1$ lies in $\mathcal{W}$.

\item $w=a_{1}a_{2}\cdots a_{\ell-1}a_{\ell}\in\mathcal{W}\Rightarrow
w_{Left}=a_{1}a_{2}\cdots a_{\ell-1}\in\mathcal{W}$, and

\item $w=a_{1}a_{2}\cdots a_{\ell-1}a_{\ell}\in\mathcal{W}\Rightarrow
w_{Right}=a_{2}\cdots a_{\ell-1}a_{\ell}\in\mathcal{W}$
\end{itemize}

\noindent Given an epimorphism $\nu:F\longrightarrow G$ of the free group $F$
onto a group $G$, a \textbf{2-sided Schreier transversal} $\tau
:G\longrightarrow F$ for $\nu$ is a transversal of $\nu$ for which there
exists a 2-sided Schreier system such that $\tau\left(  G\right)
=\mathcal{W}$. \ A 2-sided Schreier transversal is said to be \textbf{minimal}
provided the length of each word $w$ is less than or equal to the length of
each reduced word in the coset $wKer\left(  \nu\right)  =Ker\left(
\nu\right)  w$, where $Ker\left(  \nu\right)  $ denotes the kernel of the
epimorpism $\nu$.
\end{definition}

\bigskip

The wandering Shor algorithm found in section 7 suggests that a correct
generalization of the Shor transversal $n\operatorname{mod}N\longmapsto n$
$(0\leq n<Q)$ must at least have the property that it is a minimal 2-sided
Schreier transversal. \ Whatetever other additional properties this
generalization must have is simply not clear. \ 

\bigskip

In \cite{Lomonaco12}, we construct and investigate a number of different QHS
algorithms on free groups that arise from the application of various
additional conditions imposed upon the minimal 2-sided Schreier transversal
requirement. \ In this section, we only give a descriptive sketch of the
simplest of these algorithms, i.e., a QHS algorithm on free groups with only
the minimal 2-sided Schreier transversal requirement imposed.

\bigskip

Let $F=F\left(  x_{1},x_{2},\ldots,x_{n}\right)  $ be the free group of finite
rank $n$ with free basis $X=\left\{  x_{1},x_{2},\ldots,x_{n}\right\}  $, and
let $\varphi:F\longrightarrow S$ be an HSP on the free group $F$. \ We assume
that the hidden subgroup $K$ is normal and of finite index in $F$. \ (Please
note that $K=Ker\left(  \varphi\right)  =\varphi^{-1}\varphi\left(  1\right)
$ .)

\bigskip

\begin{itemize}
\item Choose a finite group probe $G$ with presentation $\left(  x_{1}%
,x_{2},\ldots,x_{n}:r_{1},r_{2},\ldots,r_{m}\right)  _{\nu}$, where the
subscript $\nu$ denotes the epimorphism $\nu:F\longrightarrow G$ induced by
the map $x_{j}\longmapsto x_{j}Cons\left(  r_{2},\ldots,r_{m}\right)  $. \ 

\item Choose a minimal 2-sided Schreier transversal $\tau:G\longrightarrow F$
of the epimorphism $\nu:F\longrightarrow G$.

\item Finally, construct the push
\[
\widetilde{\varphi}=Push\left(  \varphi\right)  =\varphi\circ\tau
:G\longrightarrow S
\]

\end{itemize}

\bigskip

Our generalized Shor algorithm for the free group $F$ consists of the
following steps:

\bigskip

\begin{itemize}
\item[\textbf{Step 1.}] Call \textsc{QRand}$\left(  \widetilde{\varphi
}\right)  $ to produce a word $s_{j}^{\prime}$ in $F$ close to a word $s_{j}$
lying in $\varphi^{-1}\varphi\left(  1\right)  $.

\item[\textbf{Step 2.}] With input $s_{j}^{\prime}$, use a polytime classical
algorithm to determine $s_{j}$. \ (See \cite{Lomonaco12}.)

\item[\textbf{Step 3.}] Repeat Steps 1 and 2 until enough relators $s_{j}$'s
are found to produce a presentation
\[
\left(  x_{1},x_{2},\ldots,x_{n}:s_{1},s_{2},\ldots,s_{\ell}\right)
\]
of the hidden subgroup $F/K$, then output the presentation $\left(
x_{1},x_{2},\ldots,x_{n}:s_{1},s_{2},\ldots,s_{\ell}\right)  $, and
\textsc{Stop}.
\end{itemize}

\bigskip

Obviously, much more needs to be said. For, example, we have not explained how
one chooses the relators $r_{j}$ so that $G=\left(  x_{1},x_{2},\ldots
,x_{n}:r_{1},r_{2},\ldots,r_{m}\right)  $ is a good group probe. \ Moreover,
we have not explained what classical algorithm is used to transform the words
$s_{j}^{\prime}$ into the relators $s_{j}$. \ For more details, we refer the
reader to \cite{Lomonaco12}.

\bigskip

\section{Is Grover's algorithm a QHS algorithm?}

\bigskip

In this section, our objective is to factor Grover's algorithm into the QHS
primitives developed in the previous sections of this paper. \ As a result, we
will show that Grover's algorithm is more closely related to Shor's algorithm
than one might at first expect. \ In particular, we will show that Grover's
algorithm is a QHS algorithm in the sense that it solves an HSP $\varphi
:S_{N}\longrightarrow S$, which we will refer to as the \textbf{Grover HSP.}
\ \ However, we will then show that the standard QHS algorithm for this HSP
cannot possibly find a solution.

We begin with a question:

\bigskip

\begin{center}
\textit{Does Grover's algorithm have symmetries that we can exploit?}
\end{center}

\bigskip

The problem solved by Grover's algorithm \cite{Lomonaco3}, \cite{Grover1},
\cite{Grover2}, \cite{Grover3} is that of finding an unknown integer label
$j_{0}$ in an unstructured database with items labeled by the integers:%

\[
0,1,2,\ldots,j_{0},\ldots,N-1=2^{n}-1\text{ \ ,}%
\]
given the oracle%
\[
f\left(  j\right)  =\left\{
\begin{array}
[c]{ll}%
1 & \text{if \ }j=j_{0}\\
0 & \text{otherwise}%
\end{array}
\right.
\]

\bigskip

Let $\mathcal{H}$ be the Hilbert space with orthonormal basis $\left\vert
0\right\rangle ,\left\vert 1\right\rangle ,\left\vert 2\right\rangle
,\ldots,\left\vert N-1\right\rangle $. \ Grover's oracle is essentially given
by the unitary transformation%
\begin{gather*}
I_{\left\vert j_{0}\right\rangle }:\mathcal{H}\;\;\longrightarrow
\qquad\;\mathcal{H}\\
\quad\;\;\left\vert j\right\rangle \;\;\longmapsto\;\;\left(  -1\right)
^{f(j)}\left\vert j\right\rangle
\end{gather*}
where $I_{\left\vert j_{0}\right\rangle }=I-2\left\vert j_{0}\right\rangle
\left\langle j_{0}\right\vert $ is inversion in the hyperplane orthogonal to
$\left\vert j\right\rangle $. Let $W$ denote the Hadamard transformation on
the Hilbert space $H$. Then Grover's algorithm is as follows:

\bigskip

\begin{itemize}
\item[\textbf{Step 0.}] (Initialization)%
\begin{gather*}
\left\vert \psi\right\rangle \;\;\longleftarrow\;\;W\left\vert 0\right\rangle
=\dfrac{1}{\sqrt{N}}\sum\limits_{j=0}^{N-1}\left\vert j\right\rangle \\
k\quad\;\,\longleftarrow\;\;0
\end{gather*}

\item[\textbf{Step 1.}] Loop until $k\approx\pi\sqrt{N}/4$%
\begin{gather*}
\left\vert \psi\right\rangle \;\;\longleftarrow\;\;Q\left\vert \psi
\right\rangle =-WI_{\left\vert 0\right\rangle }WI_{\left\vert j_{0}%
\right\rangle }\left\vert \mathbf{\psi}\right\rangle \\
k\quad\;\,\longleftarrow\;\;k+1
\end{gather*}

\item[\textbf{Step 2.}] Measure $\left\vert \psi\right\rangle $ with respect
to the standard basis%
\[
\left\vert 0\right\rangle ,\left\vert 1\right\rangle ,\left\vert
2\right\rangle ,\ldots,\left\vert N-1\right\rangle
\]
to obtain the unknown state $\left\vert j_{0}\right\rangle $ with%
\[
Prob\geq1-\dfrac{1}{N}%
\]

\end{itemize}

\bigskip

But where is the hidden symmetry in Grover's algorithm?

\bigskip

Let $S_{N}$ be the symmetric group on the symbols $0,1,2,\ldots,N-1$. \ Then
Grover's algorithm is invariant under the \textbf{hidden subgroup
}$Stab_{j_{0}}=\left\{  g\in S_{N}:g\left(  j_{0}\right)  =j_{0}\right\}
\subset S_{N}$,called the \textbf{stabilizer subgroup} for $j_{0}$, i.e.,
Grover's algorithm is invariant under the group action%
\begin{gather*}
\;\;Stab_{j_{0}}\times H\quad\;\;\longrightarrow\qquad\;\;H\\
\left(  g,\sum\nolimits_{j=0}^{N-1}a_{j}\left\vert j\right\rangle \right)
\;\;\,\longmapsto\;\;\sum\nolimits_{j=0}^{N-1}a_{j}\left\vert g\left(
j\right)  \right\rangle
\end{gather*}

\bigskip

Moreover, if we know the hidden subgroup $Stab_{j_{0}}$, then we know $j_{0}$,
and vice versa. In other words, the problem of finding the unknown label
$j_{0}$ is informationally the same as the problem of finding the hidden
subgroup $Stab_{j_{0}}$.

\bigskip

Let $\left(  ij\right)  \in S_{N}$ denote the permutation that interchanges
integers $i$ and $j$, and leaves all other integers fixed. Thus, $\left(
ij\right)  $is a transposition if $i\neq j$, and the identity permutation $1$
if $i=j$.

\bigskip

\begin{proposition}
The set $\left\{  \left(  0j_{0}\right)  ,\left(  1j_{0}\right)  ,\left(
2j_{0}\right)  ,\ldots,\left(  \left(  N-1\right)  j_{0}\right)  \right\}  $
is a complete set of distinct coset representatives for the hidden subgroup
$Stab_{j_{0}}$ of $S_{N}$, i.e., the coset space $S_{N}/Stab_{j_{0}}$is given
by the following complete set of distinct cosets:%
\[
S_{N}/Stab_{j_{0}}=\left\{  \left(  0j_{0}\right)  Stab_{j_{0}},\left(
1j_{0}\right)  Stab_{j_{0}},\left(  2j_{0}\right)  Stab_{j_{0}},\ldots,\left(
\left(  N-1\right)  j_{0}\right)  Stab_{j_{0}}\right\}
\]

\end{proposition}

\bigskip

We can now see that Grover's algorithm is a hidden subgroup algorithm in the
sense that it is a quantum algorithm which solves the following hidden
subgroup problem:

\bigskip

\noindent\textbf{Grover's Hidden Subgroup Problem.} \textit{Let }%
$\varphi:S_{N}\longrightarrow S$\textit{ be a map from the symmetric group
}$S_{N}$\textit{ to a set }$S=\left\{  0,1,2,\ldots,N-1\right\}  $\textit{
with hidden subgroup structure given by the commutative diagram}%
\[%
\begin{array}
[c]{ccc}%
S_{N} & \longrightarrow & \quad\;\;S\\
\nu_{j_{0}}\searrow &  & \nearrow\iota\\
& S_{N}/Stab_{j_{0}} &
\end{array}
\text{ \ ,}%
\]
\textit{where }$\nu_{j_{0}}:S_{N}\longrightarrow S_{N}/Stab_{j_{0}}$\textit{
is the natural surjection of }$S_{N}$\textit{ on to the coset space }%
$S_{N}/Stab_{j_{0}}$\textit{, and where}%

\[%
\begin{array}
[c]{ccc}%
\iota:\;S_{N}/Stab_{j_{0}} & \longrightarrow & S\\
\quad\left(  jj_{0}\right)  Stab_{j_{0}} & \longmapsto & j
\end{array}
\]
\textit{is the unknown relabeling (bijection) of the coset space }%
$S_{N}/Stab_{j_{0}}$\textit{onto the set }$S$\textit{. Find the hidden
subgroup }$Stab_{j_{0}}$\textit{with bounded probability of error.}

\bigskip

Now let us compare Shor's algorithm with Grover's.

\bigskip

From section 6, we know that Shor's algorithm \cite{Lomonaco2},
\cite{Lomonaco4}, \cite{Shor1}, \cite{Shor2} solves the hidden subgroup
problem $\varphi:\mathbb{Z}\longrightarrow\mathbb{Z}_{N}$ with hidden subgroup
structure%
\[%
\begin{array}
[c]{ccc}%
\mathbb{Z} & \longrightarrow & \quad\;\;\mathbb{Z}_{N}\\
\;\;\nu\searrow &  & \nearrow\iota\\
& \mathbb{Z}/P\mathbb{Z} &
\end{array}
\]
Moreover, as stated in section 6, Shor has created his algorithm by
pushing\footnote{See Section II.A.6 for a definition of pushing.} the above
hidden subgroup problem $\varphi:\mathbb{Z}\longrightarrow\mathbb{Z}_{N}$ to
the hidden subgroup problem $\widetilde{\varphi}:\mathbb{Z}_{Q}\longrightarrow
\mathbb{Z}_{N}$ (called Shor's oracle), where the hidden subgroup structure of
$\widetilde{\varphi}$ is given by the commutative diagram%
\[%
\begin{array}
[c]{ccc}%
Z & \longrightarrow & \quad\;\;Z_{N}\\
\qquad\;\;\alpha\searrow\nwarrow\tau &  & \nearrow\widetilde{\varphi}%
=\varphi\circ\tau\\
& Z_{Q} &
\end{array}
\text{ \ ,}%
\]
where $\alpha$ is the natural epimorphism of $\mathbb{Z}$ onto $\mathbb{Z}%
_{Q}$, and where $\tau$ is Shor's chosen transversal for the epimorphism
$\alpha$.

\bigskip

Surprisingly, Grover's algorithm, viewed as an algorithm that solves the
Grover hidden subgroup problem, is very similar to Shor's algorithm.

\bigskip

Like Shor's algorithm, Grover's algorithm solves a hidden subgroup problem,
i.e., the Grover hidden subgroup problem $\varphi:S_{N}\longrightarrow S$ with
hidden subgroup structure%
\[%
\begin{array}
[c]{ccc}%
S_{N} & \longrightarrow & \quad\;\;S\\
\;\;\nu\searrow &  & \nearrow\iota\\
& S_{N}/Stab_{j_{0}} &
\end{array}
\text{ \ ,}%
\]
where $S=\left\{  0,1,2,\ldots,N-1\right\}  $ denotes the set resulting from
an unknown relabeling (bijection)
\[
\left(  jj_{0}\right)  Stab_{j_{0}}\longmapsto j
\]
of the coset space%
\[
S_{N}/Stab_{j_{0}}=\left\{  \left(  0j_{0}\right)  Stab_{j_{0}},\left(
1j_{0}\right)  Stab_{j_{0}},\left(  2j_{0}\right)  Stab_{j_{0}},\ldots,\left(
\left(  N-1\right)  j_{0}\right)  Stab_{j_{0}}\right\}  \text{ \ .}%
\]

\bigskip

Also, like Shor's algorithm, we can think of Grover's algorithm as one created
by pushing the Grover hidden subgroup problem $\varphi:S_{N}\longrightarrow S$
to the hidden subgroup problem $\widetilde{\varphi}:S_{N}/Stab_{j_{0}%
}\longrightarrow S$, where the pushing is defined by the following commutative
diagram%
\[%
\begin{array}
[c]{ccc}%
S_{N} & \longrightarrow & \qquad\qquad\;\;S=S_{N}/Stab_{j_{0}}\\
\qquad\;\;\alpha\searrow\nwarrow\tau &  & \nearrow\widetilde{\varphi}%
=\varphi\circ\tau\\
& S_{N}/Stab_{0} &
\end{array}
\text{ \ \ ,}%
\]
where $\alpha:S_{N}\longrightarrow S_{N}/Stab_{0}$ denotes the natural
surjection of $S_{N}$ onto the coset space $S_{N}/Stab_{0}$, and where
$\tau:S_{N}/Stab_{0}\longrightarrow S_{N}$denotes the transversal of $\alpha$
given by%
\[%
\begin{array}
[c]{ccc}%
S_{N}/Stab_{0} & \longrightarrow & S_{N}\\
\left(  j0\right)  Stab_{0} & \longmapsto & \left(  j0\right)
\end{array}
\text{ \ \ .}%
\]

\bigskip

Again also like Shor's algorithm, the map $\widetilde{\varphi}$ given by
\begin{gather*}
S_{N}/Stab_{0}\quad\longrightarrow\quad\;S_{N}/Stab_{j_{0}}=S\\
\left(  j0\right)  Stab_{0}\;\;\mapsto\;\;\left(  jj_{0}\right)  Stab_{j_{0}%
}=j
\end{gather*}
is (if $j_{0}\neq0$) actually a disguised Grover's oracle. For the map
$\widetilde{\varphi}$ can easily be shown to simply to%
\[
\widetilde{\varphi}\left(  \mathbf{(}j0\mathbf{)}Stab_{0}\right)  =\left\{
\begin{array}
[c]{ll}%
(j0)Stab_{j_{0}} & \text{if \ }j=j_{0}\\
Stab_{j_{0}} & \text{otherwise\ \ ,}%
\end{array}
\right.
\]
which is informationally the same as Grover's oracle%
\[
f\left(  j\right)  =\left\{
\begin{array}
[c]{ll}%
j & \text{if \ }j=j_{0}\\
1 & \text{otherwise}%
\end{array}
\right.
\]

\bigskip

Hence, we can conclude that Grover's algorithm is a quantum algorithm very
much like Shor's algorithm, in that it is a quantum algorithm that solves the
Grover hidden subgroup problem.

\bigskip

However, \dots, this appears to be where the similarity between Grover's and
Shor's algorithms ends. For the standard non-abelian QHS algorithm for $S_{N}$
cannot find the hidden subgroup $Stab_{j_{0}}$ for each of following two reasons:

\begin{itemize}
\item Since the subgroups $Stab_{j}$ are not normal subgroups of $S_{N}$, it
follows from the work of Hallgren et al \cite{Hallgren1}, \cite{Hallgren2}
that the standard non-abelian hidden subgroup algorithm will find the largest
normal subgroup of $S_{N}$ lying in $Stab_{j_{0}}$. But unfortunately, the
largest normal subgroup of $S_{N}$ lying in $Stab_{j}$ is the trivial subgroup
of $S_{N}$.

\item The subgroups $Stab_{0},Stab_{1},\ldots,Stab_{N-1}$ are mutually
conjugate subgroups of $S_{N}$. \ Moreover, one can not hope to use this QHS
approach to Grover's algorithm to find a faster quantum algorithm. For Zalka
\cite{Zalka1} has shown that Grover's algorithm is optimal.
\end{itemize}

\bigskip

The arguments given above suggest that Grover's and Shor's algorithms are more
closely related that one might at first expect. Although the standard
non-abelian QHS algorithm on $S_{N}$ can not solve the Grover hidden subgroup
problem, there does remain an intriguing question:

\bigskip

\noindent\textbf{Question.} \textit{Is there some modification of (or
extension of) the standard QHS algorithm on the symmetric group }$S_{N}%
$\textit{ that actually solves Grover's hidden subgroup problem?}

\bigskip

For a more in-depth discussion of the results found in this section, we refer
the reader to \cite{Lomonaco11}.

\bigskip

\section{Beyond QHS algorithms: The suggestions of a meta-scheme for creating
new quantum algorithms}

\bigskip

In this paper, we have decomposed Shor's quantum factoring algorithm into
primitives, generalized these primitives, and then reassembled them into a
wealth of new QHS algorithms. \ But as the results found in the previous
section suggest, this list of quantum algorithmic primitives is far from
complete. \ This is expressed by the following question:

\bigskip

\begin{center}
\textit{Where can we find more algorithmic primitives to create a more well
rounded toolkit for quantum algorithmic development?}
\end{center}

\bigskip

The previous section suggests that indeed all quantum algorithms may well be
hidden subgroup algorithms in the sense that they all find hidden symmetries,
i.e., hidden subgroups. \ This is suggestive of the following meta-procedure
for quantum algorithm development:

\begin{itemize}
\item[\textbf{Meta-Step 1.}] Explicitly state the problem to be solved.

\item[\textbf{Meta-Step 2.}] Rephrase the problem as a hidden symmetry problem.

\item[\textbf{Meta-Step 3.}] Create a quantum algorithm to find the hidden symmetry.
\end{itemize}

\bigskip

\begin{center}
\textit{Can this meta-procedure be made more explicit?}
\end{center}

\bigskip

Perhaps some reader to this paper will be able to answer this question.

\bigskip

\section{Acknowledgement}

\bigskip

This work is partially supported by the Defense Advanced Research Projects
Agency (DARPA) and Air Forche Research Laboratory, Air Force Materiel Command,
USAF, under agreement number F30602-01-2-0522. The U.S. Government is
authorized to reproduce and distribute reprints for Governmental purposes
notwithstanding any copyright annotation thereon. \ This work also partially
supported by the Institute for Scientific Interchange (ISI), Torino, the
National Institute of Standards and Technology (NIST), the Mathematical
Sciences Research Institute (MSRI), the Isaac Newton Institute for
Mathematical Sciences, and the L-O-O-P fund.

\end{document}